# Radiatively Cooled Binary Mass Transfer: Flow Structure, Luminosities, and L2 Outflows Across Mass Transfer Rates

Peter Scherbak[1], Wenbin Lu[2], and Jim Fuller[1]

[1] TAPIR, California Institute of Technology, Pasadena, CA 91125, USA

[2] Departments of Astronomy and Theoretical Astrophysics Center, UC Berkeley, Berkeley, CA 94720, USA

## Abstract

High rates of stable mass transfer (MT) occur for some binary star systems, resulting in luminous transients and circumbinary outflows (CBOs). We perform hydrodynamical simulations of a $10\,M_\odot$ donor star and a $5\,M_\odot$ point mass accretor, incorporating approximate effects of radiative cooling. By varying the orbital separation of the system, we probe MT rates between $10^{-5}$ and $10^{-1} M_\odot\,\mathrm{yr}^{-1}$. Mass flows from the donor into a disk around the accretor, with significant equatorially concentrated outflows through the outer Lagrange point L2 occurring for MT rates $\gtrsim 10^{-3} M_\odot\,\mathrm{yr}^{-1}$, while the MT remains mostly conservative for lower MT rates. In all cases, any outflowing gas approximately carries the specific angular momentum of L2. The gas cooling luminosity $L$ and temperature increases with MT rate, with $L \sim 10^5 L_\odot$ and $T \sim 10^4$ K for simulations featuring the strongest outflows, with contributions from both the CBO and the accretor's disk . The most luminous transients associated with mass outflows will be rare due to the high MT rate requirement, but generate significant optical emission from both near the accretor and the CBO.

## 1. Introduction

Mass transfer (MT) in binary star systems is common, especially for binaries containing massive stars (Sana et al. 2012). Episodes of MT precede the formation of gravitational wave sources (Postnov & Yungelson 2014; Korol et al. 2022; LIGO Scientific Collaboration et al. 2023) and a large fraction of core-collapse supernova (Schneider et al. 2021). A major uncertainty related to MT is the degree to which it is conservative (Huang 1963; Mink et al. 2007), where fully conservative MT results in the accreting star accepting all material from the donor. Non-conservative MT, which occurs in contexts including super-Eddington MT onto a compact object (e.g., Toyouchi et al. 2024), spin-up of accretors to critical rotation (e.g., Lechien et al. 2025), and thermal-timescale MT (e.g., Shao & Li 2016; Pavlovskii et al. 2017; Klencki et al. 2022), results in material leaving the binary system, extracting angular momentum (AM). A consequent uncertainty is the amount of AM carried by the escaping mass.

If mass escapes by flowing through the outer Lagrange point, L2, which is a likely outcome in cases including an evolved massive star and stellar companion (Podsiadlowski et al. 1992), then its specific AM may be comparable to that of the L2 point (e.g., Huang 1963; MacLeod et al. 2018a). In studies of stable MT that can precede the formation of binary black holes (BBHs), Marchant et al. (2021) and Picco et al. (2024) noted that L2 mass loss may occur and that the amount of specific AM lost is a significant uncertainty that will impact the observed BBH population.

L2 mass loss has been predicted for high rates of MT ($\gtrsim 10^{-4} M_\odot\,\mathrm{yr}^{-1}$), where gas in the accretion disk is expected to cool inefficiently and be energetic enough to escape through L2, forming a circumbinary outflow (CBO; Lu et al. 2023). The observational appearance of the MT is unclear, with Lu et al. (2023) predicting that photons from the accretion disk will be reprocessed into the infrared.

Recently, Scherbak et al. (2025) (hereafter Paper I) performed hydrodynamical simulations of rapid, stable MT that adopted an adiabatic assumption and therefore neglected cooling. Paper I found that material indeed flowed out through L2, carrying AM similar to that of L2, and with a relatively small average radial velocity (10%–20% the orbital velocity). At the outer boundary of those simulations, the material remained marginally bound to the binary, with slightly negative specific energy. This meant it was unclear if L2 mass loss in a binary could be a mechanism behind pre-supernova mass loss and extended circumstellar material that commonly interacts with supernova ejecta (e.g., Taddia et al. 2013; Clark et al. 2020).

A major assumption in Paper I was the lack of radiative cooling, which Paper I estimated to become important at moderate MT rates $\lesssim 10^{-2} M_\odot$ yr$^{-1}$. While the MT in Paper I was highly non-conservative when the simulations reached a steady-state, Lu et al. (2023) has predicted that the degree of non-conservativeness should increase with the MT rate, further suggesting that behavior could differ at lower MT rates. In this paper, we extend the work of Paper I by incorporating radiative cooling into hydrodynamical simulations, thereby expanding our simulations to cover lower and more common MT rates, where L2 mass loss may occur differently or not occur at all. In addition, the presence of cooling will result in more realistic gas temperatures, and allow for calculations of the observational appearance of the the accretor's disk and CBOs. Because we do not capture viscous accretion processes, we do not explicitly refer to an accretion disk around the accretor, but instead use terms such as "the accretor's disk," or "the disk around the accretor."

Observationally, L2 mass loss appears to be occurring in binaries SS 433 (Blundell et al. 2001; Cherepashchuk et al. 2018) and W Serpentis (Shepard et al. 2024). In the case of SS 433, radio/infrared images (Blundell et al. 2001; Bowler 2010, 2011a, 2011b; Perez & Blundell 2010) suggest the presence of outflowing circumbinary material. Infrared images of W Serpentis (Davidge 2023) also find dust formation extending away from the central binary, suggesting the formation of a circumbinary disk that outflows through the outer Lagrange point on the side of the accretor (Shepard et al. 2024). The red transient V1309 Sco may also have undergone L2 mass loss, which would explain several peculiar features of its light curve (Pejcha 2014).

Numerous works have investigated mass loss through the outer Lagrange point (summarized in Paper I) but only a relatively small subset have incorporated cooling and investigated various MT rates. Pejcha et al. (2016) performed smoothed particle hydrodynamic (SPH) simulations of L2 mass-loss with radiative cooling and heating, but they initialized their material near L2 without including the donor star or accretor in their domain. Nazarenko et al. (2005) modeled MT rates between $\sim 10^{-7}$ and $10^{-5}$ $M_\odot$ yr$^{-1}$ in Algol-type binaries, noting that gas in the accretion disk can be transferred to the vicinity of L2/L3 and form an outflow, but did not discuss the fraction of material lost from the accretion disk. Mohamed & Podsiadlowski (2012) studied MT in symbiotic binaries undergoing wind Roche lobe overflow, finding that some of the wind is deflected away from the accretor and escapes through L2. Booth et al. (2016) extended this work to symbiotic binaries undergoing mass loss through a combination of winds and a Roche lobe overflow stream, incorporating a sophisticated radiative cooling model. They also found that part of the wind is channeled near L2 and escapes. Both Mohamed & Podsiadlowski (2012) and Booth et al. (2016) focused on MT rates $\sim 10^{-6}$ $M_\odot$ yr$^{-1}$, whereas we focus on higher MT rates. Although they did not include cooling, Bobrick et al. (2017) investigated realistic MT rates between a WD and a neutron star through SPH simulations, finding that mass is lost through an accretion disk wind, forming a "common envelope" around both stars. The AM carried by these winds is greater than the specific orbital AM of the neutron star, but much less than the specific AM of L2. In contrast to these works, we focus on the approximately steady-state behavior of stable MT where we track the outflow of gas, originating in the disk around the accretor, from the binary, and we vary MT rates to investigate its changing effects.

In Section 2 we discuss the numerical setup of our simulations, including the grid setup and our radiative cooling model, as well as the MT rates we probe and the astrophysical systems they may apply to. Section 3.1 characterizes the gas flow and degree of conservativeness for different MT rates, Section 3.2 the temperature, luminosity and observational appearance of the MT episodes we model, and Section 3.3 the properties of any outflows through L2, with a comparison to previous work. We conclude in Section 4.

## 2. Simulation Setup

We perform Newtonian hydrodynamic simulations using the PLUTO code (Mignone et al. 2007, 2012), considering MT between a donor star with core mass $M_d$ and an accretor with mass $M_a$, separated by a distance $a$ in a fixed circular orbit. We solve the conservation equations of mass, momentum, and energy without any explicit viscosity, but including cooling due to radiative losses (Section 2.1). Our setup, including grid spacing, initial hydrostatic structure of the donor star, and injection of heat inside the donor star, is discussed in more detail in Paper I.

Our simulations are set in spherical coordinates $r$, $\theta$, $\phi$, centered at the donor star, in the co-rotating frame of the binary. The grid radius $r$ extends from inside the donor's envelope, not including the core, out to $5a$, with logarithmically spaced cells such that $\delta r/r = 0.01$ everywhere. The outer boundary condition is outflow, enforcing zero gradients of fluid quantities (see PLUTO documentation for more details), except for the radial velocity $v_r$, which is also enforced to be $v_r \geqslant 0$. With this "diode" boundary condition, we do not see any instabilities or reflection at the outer boundary. $\theta$ extends from 0 to $\pi/2$, with a reflective boundary condition, so that we only simulate the top hemisphere of the binary, while $\phi$ wraps around from 0 to $2\pi$. $\theta$ and $\phi$ are spaced uniformly with size $\delta r/r$ such that cells are approximately cubic. We refer to values in code units when we non-dimensionalize using $G = M_{\rm tot} = M_d + M_a = a = 1$, and time $t$ is given in units of the binary orbital period $P_{\rm orb} = 2\pi\sqrt{a^3/G(M_d + M_a)}$.

We use the Roche_tidal_equilibrium code of Lu (2025) to help construct our donor star setup, which co-rotates with the

**Table 1**
Grid of 3D Simulations, Labeled "Highest" "High" "Mid" or "Low" by their MT Rate in Physical Units ($\dot{M}_{\rm phys}$)

| Simulation | $\dot{M}_{\rm code}$ | $a$ (au) | $P_{\rm orb}$ (days) | $M_{\rm tot}^{3/2}G^{1/2}a^{-3/2}$ ($M_\odot$ yr $^{-1}$) | $\dot{M}_{\rm phys}$ ($M_\odot$ yr $^{-1}$) |
|---|---|---|---|---|---|
| `highest_mdot` | 1.3e-5 | 0.1 | 3 | 1.1e4 | 1.5e-1 |
| `high_mdot` | 1.1e-5 | 0.3 | 15 | 2.2e3 | 2.3e-2 |
| `mid_mdot` | 0.7e-5 | 1 | 100 | 350 | 2.2e-3 |
| `low_mdot` | 0.4e-5 | 10 | 2970 | 5 | 2.0e-5 |

**Note.** The free parameter is the orbital separation $a$, while $M_{\rm tot}$ is fixed at 15 $M_\odot$. The corresponding orbital period $P_{\rm orb}$ is calculated via Kepler's Third Law. The factor $M_{\rm tot}G^{1/2}a^{-3/2}$ is used to convert from $\dot{M}_{\rm code}$ to $\dot{M}_{\rm phys}$ (see Equation (2)), where both values are approximate steady-state values.

binary. The density $\rho$ and pressure $P$ inside the donor star are initialized using the conditions: 1. $P = K\rho^\gamma$, for some entropy constant $K$ and polytropic index $\gamma$; 2. the star is initially in hydrostatic equilibrium in the Roche potential and underfills its Roche lobe. Outside the star, $\rho$ and $P$ are initialized to spatially constant small values, corresponding to a low sound speed $c_s$. $P$ and the internal energy $u_{\rm int}$ are related by PLUTO's "ideal" equation of state (EOS),

$$P = (\gamma - 1)\rho u_{\rm int}, \quad (1)$$

where we use the same value of $\gamma$ as the polytopic index above. We inject heat inside the envelope so that it expands on a Kelvin–Helmholtz timescale $t_{\rm KH}$ much greater than the orbital dynamical timescale.

The accretor is represented by a Plummer softening potential, with softening length $\epsilon$ such that $\epsilon/a = 0.05$. The values $M_d$ and $M_a$ do not change over the course of the simulations, but the mass of the donor's envelope is reduced as MT occurs. We neglect the self-gravity of the donor's envelope, as well as the self-gravity of any subsequent flows, because the mass in the envelope is, by design, much less than $M_d$ and $M_a$.

The values controlling our setup are the same as in our fiducial simulation of Paper I, `q_0.5_mid_heat`: mass ratio $q \equiv \frac{M_a}{M_d} = 0.5$, $K = 0.578$ (code units), $\gamma = 1.4$ and an initial donor radius of about 75% the Roche radius. The total luminosity injected inside the donor is 4.0e-6 (code units) and $t_{\rm KH}$ is about 80 orbital periods. Although Paper I found that varying $q$ can impact outflow properties, we leave $q$ fixed in this work to focus on the impact of cooling, but future work should incorporate cooling and vary MT rates with different $q$ values.

Because our goal is to incorporate cooling and investigate different MT regimes, conversions are necessary to convert from code units and give physical values of temperatures, opacities, and cooling rates. This differs from the previous simulations of Paper I which assumed adiabaticity outside the donor star, did not incorporate cooling, and were therefore, in principle, scale-free. We choose $M_{\rm tot} = 15\,M_\odot$, which could correspond to a massive star plus stellar or black hole accretor (although see below a discussion of which companion is most valid given our assumptions). We vary the value of $a$ between our simulations, which effectively changes the physical MT rate between simulations via

$$\dot{M}_{\rm phys} = \left(\frac{M_{\rm tot}^{3/2}G^{1/2}}{a^{3/2}}\right)\dot{M}_{\rm code}. \quad (2)$$

By raising $a$ and keeping $M_{\rm tot}$ fixed and $\dot{M}_{\rm code}$ approximately fixed, we lower $\dot{M}_{\rm phys}$. Strictly speaking, this means that we do not sample all regimes of parameter space (e.g., low $\dot{M}_{\rm phys}$ at short orbital period). However, it is reasonable to expect that $\dot{M}_{\rm phys}$ will be the main factor in determining outflow properties, and that our results can be extended to various $a$. Similarly, setting the scale of $a$ determines the physical values of density, pressure, temperature, opacity, etc.

Table 1 summarizes the simulations that we perform, demonstrating the difference in scales used to simulate different MT rates. Note that values such as $K$ and $L_{\rm heat}$, which are constant in code units across simulations, are different in physical units for different simulations. See Section 2.2 for a comparison to astrophysical MT rates, and Figure 4 for further discussion of where our simulations fall in the parameter space of MT rate and $a$.

In our main suite of simulations, we do not include any heating terms due to the luminosity of either the donor star or the accretor. We do not include an explicit viscosity so we cannot capture viscous heating effects, nor can we resolve an inner accretion disk. We therefore do not model the potential launching of a fast wind that then emits X-ray and UV photons (Lu et al. 2023), or the formation of an advection dominated accretion flow (ADAF. Yuan & Narayan 2014; Yuan et al. 2015).

Although we are formally agnostic to the nature of $M_a$, our setup is more appropriate when the accretor is a star smaller in radius than $M_d$, but not a compact object. In such a case, there is no inner accretion disk, and our MT rates are not extremely super-Eddington when compared to the accretion flow power.

### 2.1. Cooling Prescription

We include energy losses due to radiative cooling, assuming that the photons primarily escape perpendicular to the orbital plane in the $\hat{z}$ direction. We estimate that the cooling power per volume, $\dot{E}_{\rm cool}$, is given by

$$\dot{E}_{\rm cool} = -\frac{4\rho\sigma_{\rm SB}T^4}{\frac{1}{\kappa} + \Sigma_z\tau_z} \quad (3)$$

where $\kappa$ is the opacity, $T$ the temperature of the gas, $\Sigma_z$ is the gas surface density in the $\hat{z}$ direction, and $\tau_z$ is the optical depth in the $\hat{z}$ direction. This equation is inspired by several previous works that have incorporated radiative cooling (Stamatellos et al. 2007; Forgan et al. 2009; Wilkins & Clarke 2012; Lombardi et al. 2015; Pejcha et al. 2016), but that performed SPH simulations. Of these, Pejcha et al. (2016) applied this cooling term to their SPH simulations of L2 mass loss, noting that their particles are not self-gravitating and the original formalism of Stamatellos et al. (2007), Forgan et al. (2009), Lombardi et al. (2015) to develop this formula does not strictly apply.

Equation (3) takes the correct forms in both the optically thick and optically thin limits. In the optically thin limit where $\frac{1}{\kappa} \gg \Sigma_z\tau_z$, the magnitude of the cooling reduces to

$$|\dot{E}_{\rm cool}| \approx \frac{4\rho\sigma_{\rm SB}T^4}{\frac{1}{\kappa}} \approx 4\rho\kappa\sigma_{\rm SB}T^4 \quad (4)$$

where this form represents the emitted radiation per unit volume of an optically thin gas. In the optically thick limit, where $\frac{1}{\kappa} \ll \Sigma_z\tau_z$, Equation (3) reduces to

$$|\dot{E}_{\rm cool}| \approx \frac{4\rho\sigma_{\rm SB}T^4}{\Sigma_z\tau_z} \approx \rho a T^4 \Big/ \left(\frac{H_z\rho\tau_z}{c}\right) = u_{\rm rad} \Big/ \left(\frac{H_z\tau_z}{c}\right) \quad (5)$$

where we have estimated that $\Sigma_z = H_z\rho$ for some scale height $H_z$, the form of which we will discuss shortly. This represents the radiation energy density over the diffusion timescale in the $\hat{z}$ direction, $\frac{H_z\tau_z}{c}$.

We calculate $T$, by solving the quartic equation for the total pressure as a sum of gas and radiation pressure

$$P = \frac{\rho k_b T}{\mu m_p} + \frac{a_{\rm rad}T^4}{3}, \quad (6)$$

which assumes gas and radiation are in equilibrium with the same $T$, where $k_b$ is the Boltzmann constant, $m_p$ is the proton mass, and $\mu$ is the mean molecular weight. In our simulations, $\mu \approx 0.62$, appropriate for ionized Solar-composition material, but does not take into account the formation of neutral hydrogen at low temperatures, which can occur in our low-MT rate simulations. This means the temperature calculation is not quite consistent at low $T$, but, compared to the impact of MT rate (Section 3.2), is likely a relatively small correction that will not severely impact our $T$ estimates. Equation (6) is solved at runtime at every cell, using a Newton-Raphson root solver.

The opacity $\kappa$ is estimated by interpolating tables of opacity at solar metallicity ($Z_\odot$), using tables conveniently collected in the Modules for Experiments in Stellar Astrophysics code (MESA Paxton et al. 2011, 2013, 2015, 2018, 2019; Jermyn et al. 2023). Specifically, we interpolate the table `gn93_z0.02_x0.7.data` in $\log T$ and $\log R \equiv \log\rho - 3\log T + 18$ to estimate $\log\kappa$, using `lowT_fa05_gn93_z0.02_x0.7.data` for $\log T < 3.75$. For any values outside the tables, we use the value of $\kappa$ at the outer edge of the table. The opacities we are using are Rosseland mean opacities, whereas they should technically be Planck mean opacities in the optically thin limit (see Stamatellos et al. 2007). We follow Pejcha et al. (2016) in neglecting this discrepancy, given the approximations already taken in using Equation (3) for our purposes. The opacity table we use, and where the points in the simulations fall on it, is visualized further in Appendix C.

The final two ingredients needed in calculating Equation (3) are $H_z$ and $\tau_z$. We estimate $\tau_z \approx \kappa\rho H_z$ at every point in the grid. This is a rough estimate of the local vertical optical depth, but doing a runtime integral over $z$-coordinate in spherical coordinates would be cumbersome and potentially introduce numerical artifacts. Our estimate of $\tau_z$ captures the fact that that the greatest contribution to $\tau_z$ is where the density is largest in the midplane. We estimate $H_z$ in a local fashion as well, using an estimate appropriate to a hydrostatic disk,

$$H_z = \frac{c_s}{\Omega_{\rm eff}} \quad (7)$$

where the sound speed $c_s$ is defined as $c_s^2 = \frac{\gamma P}{\rho}$ and we estimate the effective orbital frequency as

$$\Omega_{\rm eff}^2 = \frac{M_{\rm d}}{|\boldsymbol{r}|^3} + \frac{M_{\rm a}}{|\boldsymbol{r} - a\hat{x}|^3} \quad (8)$$

where $\vec{r}$ is the coordinate vector, as $M_d$ is located at the origin, and $M_a$ is located at $a\hat{x}$. This reduces to the correct limit in the disk around $M_a$ when the second term dominates, and far out in the equatorially concentrated CBO surrounding both $M_d$ and $M_a$ when $|\boldsymbol{r} - a\hat{x}| \sim |\boldsymbol{r}|$.

We include cooling in PLUTO using the `radiat.c` cooling module, where we set the right hand side of the energy equation that is solved by PLUTO. We do not allow any cooling inside the Roche lobe of the donor star, as that would be likely be a bad approximation given our cooling form is based off of a disk structure most valid surrounding $M_a$ and both stars, not for a spherical star.

Equation (3) gives the nominal form for the cooling power per volume, which acts to reduce $u_{\rm int}$. However, we also impose safeties and floors to avoid catastrophic cooling and unphysical values of $T$ or $c_s$. We set that in all grid cells, $u_{\rm int}$ cannot be reduced by more than 50% in a given timestep. Test runs showed that our results are not sensitive to this value. Additionally, we set a pressure floor and a corresponding internal energy floor, which are related via the ideal EOS (Equation (1)). The cooling is capped such that $u_{\rm int}$ is not allowed to fall below the internal energy floor. In low density regions, where $\rho < 10^{-5}$ in code units, we set the pressure floor to be proportional to $\rho$, corresponding to a low sound speed floor. In higher density regions, we set a pressure floor corresponding to a temperature floor of 2000 K, with pressure and temperature related by Equation (6).

We choose not to impose a temperature floor everywhere in the domain, because doing so in low density regions creates a high pressure and high sound speed. This leads to material seemingly having a large $u_{\rm int}$ and a spurious outflow of mass near the outer boundary. The downside to this approach is that $T$ of some low-density material can become lower than covered by our opacity tables. However, these low-density regions are unimportant for the cooling luminosity and L2 outflow. If we do not impose a floor on sound speed in low density regions, the pressure and sound speed can become extremely small, leading to unphysically high Mach numbers.

Additionally, we enforce floors on $\rho$ and $P$ within the `UserDefBoundaries` section of PLUTO's `init.c`. This is complementary to the previous discussion of floors, as those implied a limit on the cooling power, but did not enforce floors in general during runtime. The $\rho$ floor is set to be $10^{-10}$ in code units, and the $P$ floor is the same as above. More discussion of the floors is included in Appendix C.

In the `highest_mdot` simulation, numerical issues such as negative pressure occur near the pole, where the resolution is set to be low in order to avoid small timesteps. Therefore, we do not include cooling at $\theta < 0.3$ for this simulation. Because the density sharply drops near the pole and the outflow originates in the midplane, this will not affect our main results.

### 2.2. Comparison to Binary MT Rates

The physical MT rates we probe ($\dot{M}_{\rm phys}$ in Table 1) are relevant to several scenarios with massive stars in binaries. Stripped He star binaries undergoing case BB MT can reach MT rates of several $10^{-4} M_\odot\,{\rm yr}^{-1}$ for more than $10^4$ yr (Tauris et al. 2015). Massive stars expanding into the Hertzsprung gap and undergoing thermal timescale MT are expected to lose mass at rates between $10^{-3}$ and $10^{-2} M_\odot\,{\rm yr}^{-1}$ (Pavlovskii et al. 2017; Marchant et al. 2021; Klencki et al. 2022). Klencki et al. (2026) probed even higher MT rates, showing that MT in such binaries can remain stable almost to $10^{-1} M_\odot\,{\rm yr}^{-1}$, depending on mass ratio. Finally, the rapid expansion of stripped He stars in the weeks to years before undergoing a core-collapse supernova can result in MT rates above $10^{-2} M_\odot\,{\rm yr}^{-1}$, and the peak MT rate ($10^{-1} M_\odot\,{\rm yr}^{-1}$) can last for hundreds of binary orbits (Wu & Fuller 2022).

Therefore, our `high_mdot` ($10^{-2} M_\odot\,{\rm yr}^{-1}$) simulation may apply well to massive stars expanding on the Hertzsprung gap with a binary companion, whereas the `highest_mdot` simulation ($10^{-1} M_\odot\,{\rm yr}^{-1}$) applies only for brief evolutionary phases mentioned above (Wu & Fuller 2022; Klencki et al. 2026). The `mid_mdot` ($10^{-3} M_\odot\,{\rm yr}^{-1}$) and `low_mdot` ($10^{-4} M_\odot\,{\rm yr}^{-1}$) simulations cover the lower end of massive star thermal timescale MT rates, or potentially case BB MT from He star donors.

Observationally, the systems that have likely undergone L2 mass loss have high measured MT rates. SS 433 has a MT rate of about $10^{-4} M_\odot\,{\rm yr}^{-1}$ (Fabrika 2004; Cherepashchuk et al. 2018), W Serpentis a rate greater than $10^{-5} M_\odot\,{\rm yr}^{-1}$ (Gies et al. 2025) and V1309 Sco a rate that may have risen from $10^{-4} M_\odot\,{\rm yr}^{-1}$ to almost $10^{-2} M_\odot\,{\rm yr}^{-1}$ as the binary system approached a stellar merger (Pejcha 2014). Unfortunately, none of these systems have measurements for the outflow rate through L2, but they do show evidence for CBOs.

## 3. Analysis

### 3.1. Morphology and Mass Transfer Conservativeness

Figure 1 shows the density in the equatorial plane for our simulations, with the colormap scale shown in physical units. We show snapshots corresponding to about 100 orbital periods after initialization. By this time, the donor star has expanded due to heating and begun Roche lobe overflow, and the MT rates have reached a quasi-steady state.

In the `highest_mdot` simulation, most of the equatorial domain is flooded with material, although the densest outflow is centered near L2, the Lagrange point on the far side of $M_a$. Mass easily escapes the puffy disk around $M_a$ and flows out through L2 in a broad stream. In addition, the stream passing through L1, the inner Lagrange point, is quite puffy, and the boundary between donor, $M_a$'s disk and the MT stream is blurred. A small amount of material also flows out in a stream originating near L3 (the outer Lagrange point on the side of the donor star), and much of it falls back onto the L2 stream, forming shocks. Note that the simulations are in the co-rotating frame, and the outflow lags behind the binary orbit in this frame.

In the `high_mdot` simulation, a thick stream still flows out near L2, but lower density voids are also visible between the spiral arms of the CBO. The stream originating near L3 is also more diffuse.

The `mid_mdot` and `low_mdot` simulations are somewhat similar to one another, in that a relatively thin stream of material flows out of $M_a$'s disk near L2. The outflow through L3 is extremely tenuous in these cases. The boundary of $M_a$'s disk and the L1 stream, is more sharp in these simulations compared to `highest_mdot`. In the `low_mdot` simulation, the L2 outflow has a very low density, as does the outer part of $M_a$'s disk.

The same snapshots viewed edge-on, with a meriodional slice of the domain, are shown in Figure 2. The gas near $M_a$ forms a disk whose thickness increases substantially with increasing MT. Also with increasing MT rate, the opening angle $\theta$ of the CBO (measured away from the equator) increases, although the density is always largest in the equatorial plane. For the lower MT rate simulations, the outflow is difficult to see viewed edge-on because it reaches low densities and is concentrated near the equator. In the `mid_mdot` simulation, the wrapping of the spiral arm of the outflow shows up on the left side of the donor star, on the opposite of the accretor.

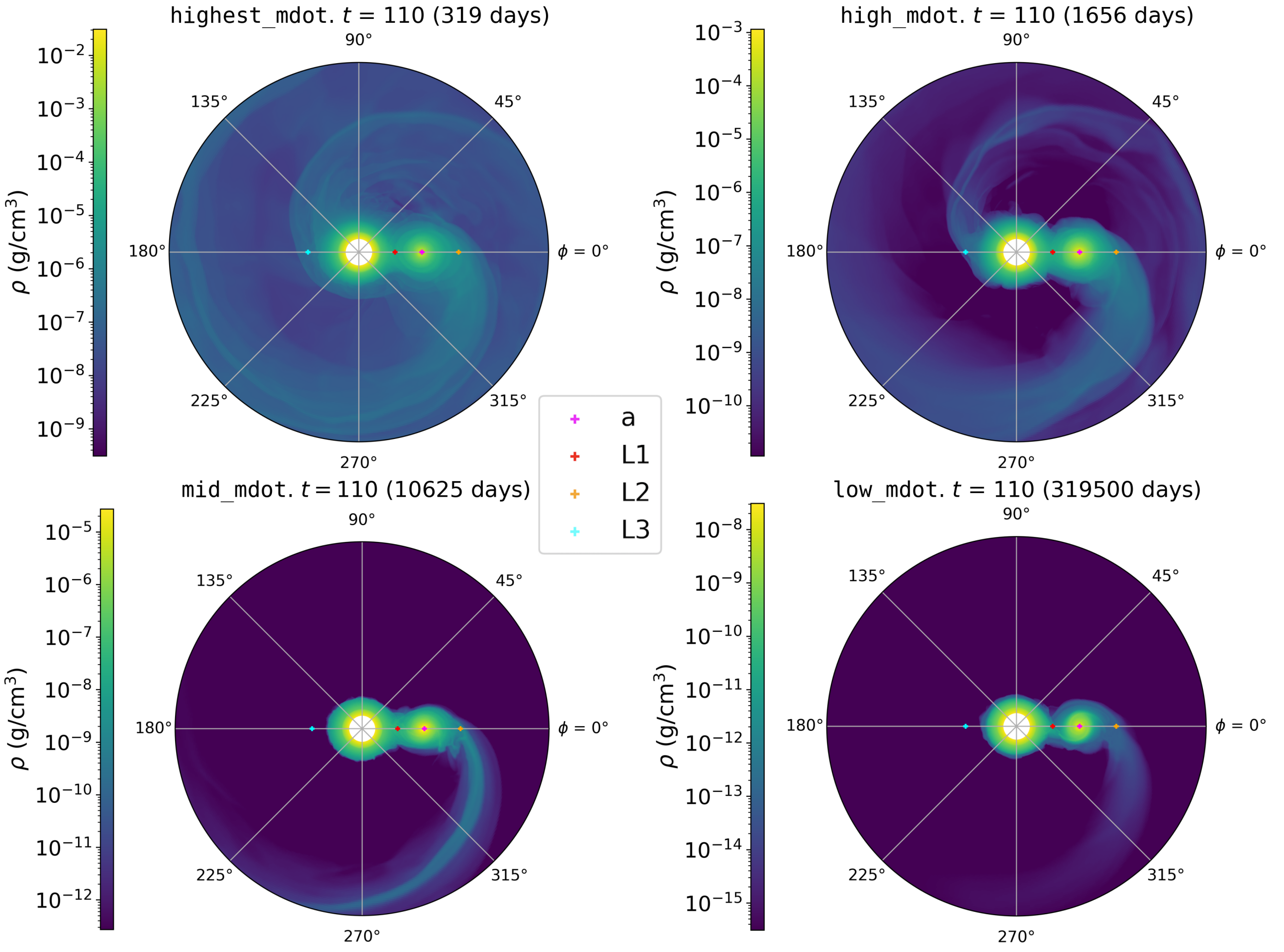

**Figure 1.** The density $\rho$ in the equatorial plane, for simulations of varying MT rate. The orbital separation $a$ and Lagrange points are labeled. The snapshots, representative of the simulations once they have reached a quasi-steady state, are labeled with $t$ in units of both orbital period and days. The full domain extends to $r = 5a$, but these plots extend to $r = 3a$ to zoom in near $M_a$. Note the different density scale in each case.

We estimate the fraction $\beta$ of transferred mass that is retained by the accretor by calculating $\dot{M}_{\rm donor}$, the MT rate from the donor star, and $\dot{M}_{\rm tot}$, the MT rate through the outer boundary, and using

$$\beta = 1 - \frac{\dot{M}_{\rm tot}}{\dot{M}_{\rm donor}}. \tag{9}$$

We note that in our simulations, material essentially either stays in the region surrounding $M_a$ or outflows near L2. There is no sink of material onto $M_a$, nor is there the opportunity for an ADAF flow to occur, which may instead expel material mostly *perpendicular* to the orbital plane in a jet or a wind (Yuan et al. 2015). Therefore, in the case of $M_a$ being a compact object, there may be a competition between outflows launched from the inner disk, and equatorial L2 outflows, which should be examined in future work.

Figure 3 plots the steady-state average of $\beta$ for our simulations. For the `highest_mdot` simulation, $\beta \approx 0$, meaning the MT is very non-conservative, which is the same conclusion of Paper I that did not include any cooling. $\beta$ monotonically increases with decreasing MT rate, meaning that the MT becomes more conservative. For the `high_mdot` simulation, $\beta \approx 0.3$–$0.5$, meaning that most of the mass still forms a CBO, but a large fraction stays near $M_a$. The other two simulations feature mostly conservative MT rates, although a small amount of material still flows out through L2. Therefore, between MT rates of $\sim 10^{-3}$ and $\sim 10^{-2} M_\odot$, there is a significant jump in the degree of non-conservativeness (top panel of Figure 3). This would constitute a threshold for highly non-conservative MT in astrophysical binaries.

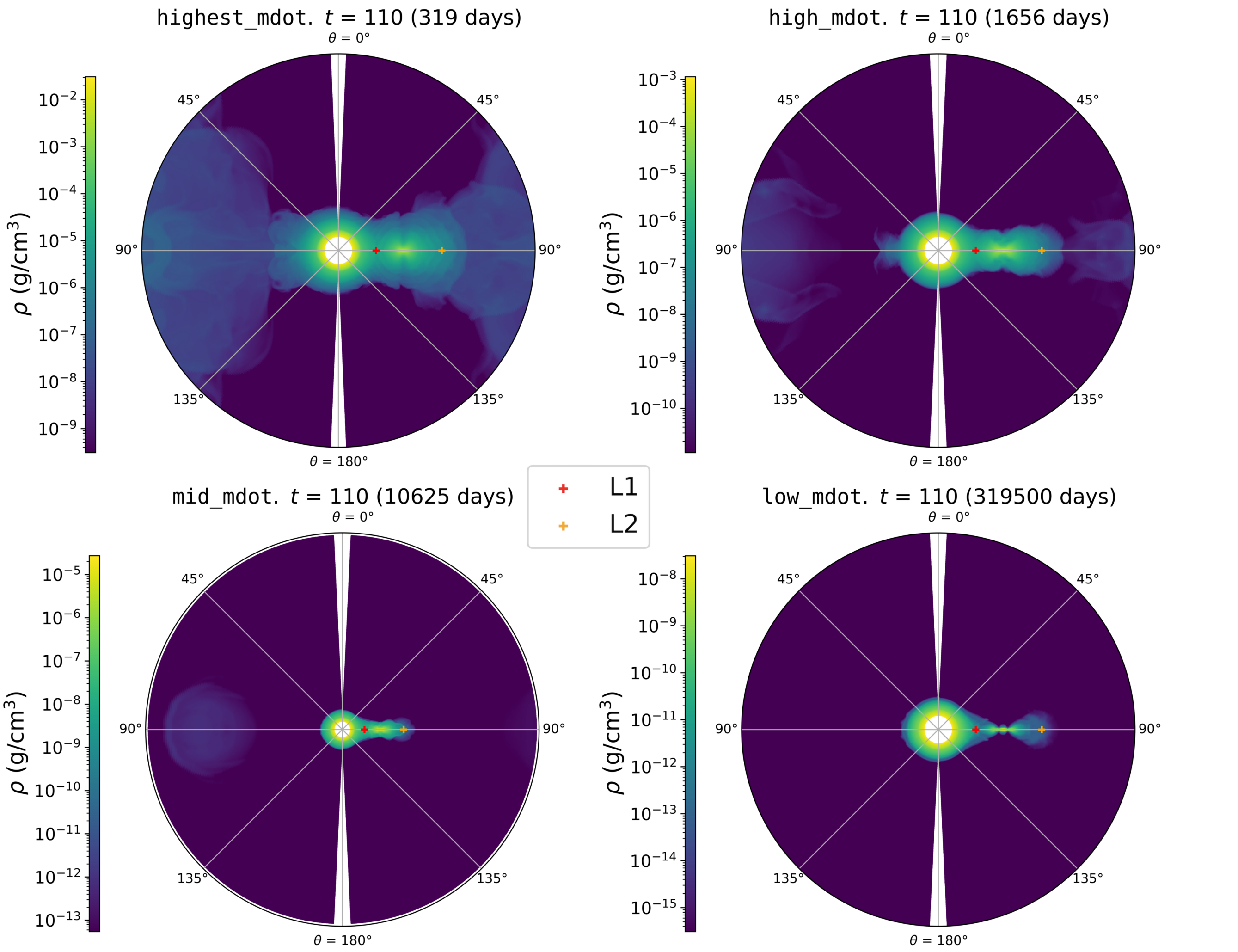

**Figure 2.** Similar to Figure 1, but plotting $\rho$ in the meridional plane lying between the donor and accretor ($\phi \approx 0$ and $\phi \approx \pi$), for simulations of varying MT rate. Plots extend to $r = 3a$, except in the case of the mid_mdot simulation, where we zoom out to $r = 5a$ to better show the outflow. We only simulate $\theta < \pi/2$, so the bottom half of each image is mirrored from the top. The white triangle near $\theta = 0$ is due to low grid resolution at the pole

In our simulations, the outflowing mass $\dot{M}_{\rm tot}$ is almost identical to the MT rate of material outward near L2, $\dot{M}_{\rm L2}$, which we calculate by integrating the mass flux over a wedge about L2, similar to the procedure in Paper I. We also estimate the MT rate near L3, $\dot{M}_{\rm L3}$, the Lagrange point on the side of the donor star. In all cases, the L2 outflow dominates, but to different degrees. For the highest_mdot and high_mdot simulations, $\dot{M}_{\rm L2} \sim 10 - 20 \times \dot{M}_{\rm L3}$, similar to our result in Paper I without any gas cooling. For the other simulations, $\dot{M}_{\rm L2} \gtrsim 10^3 \times \dot{M}_{\rm L3}$, meaning that it is extremely unfavorable for material to circle around the donor star on an equipotential that would allow it to escape near L3.

This behavior can be understood by comparing the MT rate to the Eddington-limited MT rate at the *outer* edge of the disk around $M_a$ (Lu et al. 2023). $\dot{M}_{\rm edd}$ is defined by equating the outer disk accretion luminosity to the Eddington luminosity, such that

$$\frac{GM_a|\dot{M}_{\rm donor}|}{r_d} = L_{\rm edd} = \frac{4\pi c G M_a}{\kappa} \tag{10}$$

and solving for $|\dot{M}_{\rm donor}|$, where $r_d$ is the radius near the outer edge of $M_a$'s disk. Therefore,

$$\dot{M}_{\rm edd} = \frac{4\pi c r_d}{\kappa}. \tag{11}$$

For our simulations, we estimate $\dot{M}_{\rm edd}$ by finding the density-weighted average of $\frac{4\pi c r_d}{\kappa}$ at $r_d$ values (in units of $a$) of 0.2, 0.25, and 0.3 around $M_a$ and inside $M_a$'s Roche lobe. Using lower values $r_d$ mildly changes $\dot{M}_{\rm edd}$ but not enough to change the trend in Figure 3. From the bottom panel of Figure 3, the transition between conservative and significant non-conservative MT approximately occurs when the MT rate rises above $\dot{M}_{\rm edd}$.

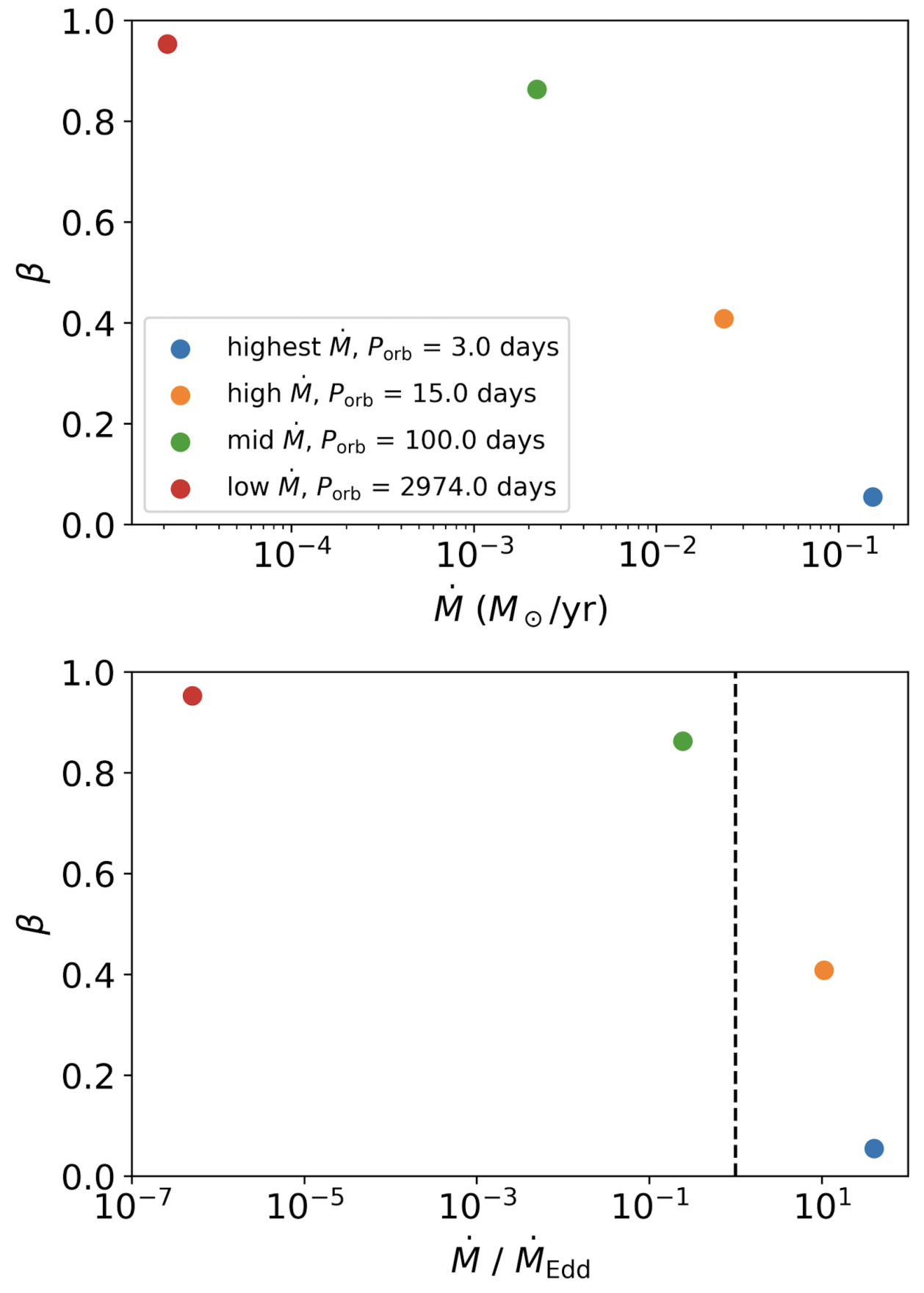

**Figure 3.** Top panel: the fraction of material retained in the disk around $M_a$, $\beta$, versus the MT rate $\dot{M}_{\rm donor}$ for our simulations. Bottom panel: the same quantity, plotted versus $\dot{M}_{\rm donor}$ in units of the Eddington MT rate (Equation (11)). The vertical dashed line shows where $\dot{M}_{\rm donor} = \dot{M}_{\rm edd}$. The values shown are time-averaged once the simulations reach a quasi-steady state, at $t$ greater than $\sim$50 (orbital periods).

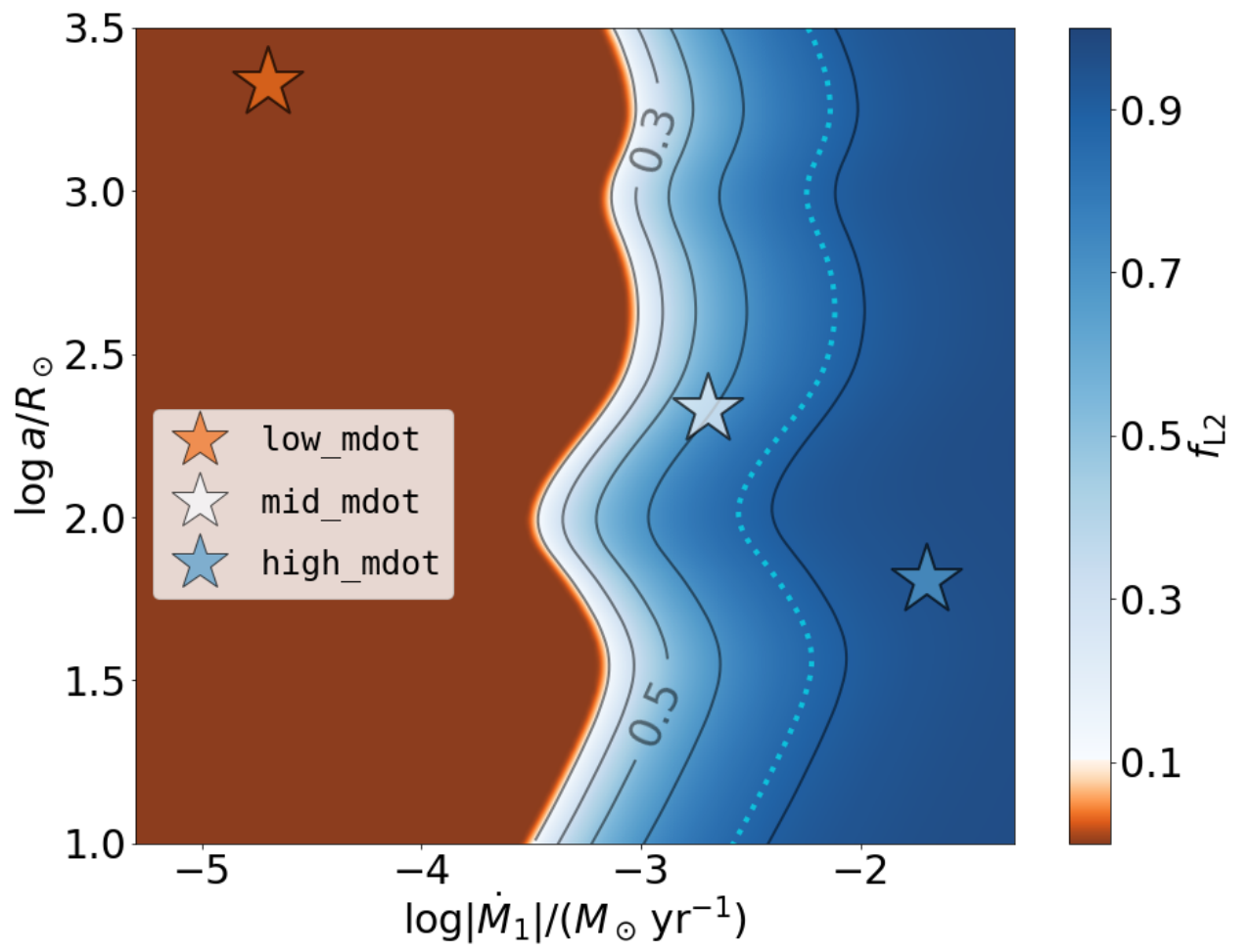

**Figure 4.** The predicted fraction of material escaping through L2, $f_{\rm L2}$, for various $a$ and MT rates. This follows the figures of Lu et al. (2023), but adapted for a 10 $M_\odot$ donor and 5 $M_\odot$ accretor. Solid lines show contours of constant $f_{\rm L2}$, and the cyan dotted line shows where $\dot{M}_{\rm donor} = \dot{M}_{\rm edd}$. Stars show where our simulations lie in this parameter space, with the internal color corresponding to their $f_{\rm L2}$ value. The `highest_mdot` simulation is not shown for convenience, but would lie outside the right border of this plot.

Figure 4 shows a comparison of our simulations to the semianalytic predictions of Lu et al. (2023), by plotting $f_{\rm L2}$, the fraction of transferred mass leaving the L2 point. Note that $f_{\rm L2}$ for our simulations is simply equal to $1 - \beta$. The colormap of Figure 4 is generated using code[3] accompanying Lu et al. (2023) for the binary masses we simulate, and assuming a hydrogen rich gas of solar metallicity.

We find that $f_{\rm L2}$ is about 0.6 for the `high_mdot` simulation and about 0.1 for the `mid_mdot` simulation, but Lu et al. (2023) predicts values of $>$0.9 and about 0.65, respectively. For our more extreme simulations, `low_mdot` and `highest_mdot`, $f_{\rm L2}$ approaches 0 and 1, in general agreement with Lu et al. (2023).

[3] https://github.com/wenbinlu/L2massloss

In other words, our simulations imply that the MT threshold where MT becomes strongly non-conservative is higher than predicted by Lu et al. (2023), by a factor of $\sim$2–5. More densely sampled simulations would help determine the differences throughout the entire parameter space (e.g., if the "wriggles" of Figure 4, caused by opacity variations, are robust). The reason for the difference is unclear, but may be due to geometric factors of order unity, our approximate treatment of gas cooling, or the lack of viscous accretion power in our simulations.

### *3.2. Observational Appearance: Temperatures and Luminosities*

We calculate the cooling luminosity of the disk around $M_a$ by integrating the power per unit volume within a sphere of radius $0.3a$, with approximates the Roche lobe of $M_a$, i.e.,

$$L_{\rm cool,disk} = \int_{|\mathbf{r} - a\hat{x}| \leqslant 0.3a} \dot{E}_{\rm cool}\, dV \tag{12}$$

and the cooling luminosity of all gas in the domain by integrating over the entire spherical domain

$$L_{\rm cool,tot} = \int_{\rm entire\ domain} \dot{E}_{\rm cool}\, dV. \tag{13}$$

Note that there is no cooling inside the Roche lobe of the donor $M_d$.

Figure 5 plots the values of luminosity, in erg s$^{-1}$, versus time for our different simulations. Because of the vastly different orbital timescales between our simulation, time is plotted in orbital periods. Unsurprisingly, the luminosity increases with increasing MT rate, although note from Appendix B that for the `highest_mdot` simulation that the gas cooling timescale is long and it is more difficult for photons to escape. For the `mid_mdot` and `low_mdot` simulations, all the cooling power comes from the disk around $M_a$, with almost nothing from the CBO. In the `highest_mdot` case, including the full domain approximately doubles the overall luminosity, so the cooling of the outflow is quite important. This is because most of the equatorial domain is flooded with material (top panel of Figure 1). The `high_mdot` simulation shows intermediate behavior, where the cooling power of the outflow makes a ∼20% increase to the overall cooling power.

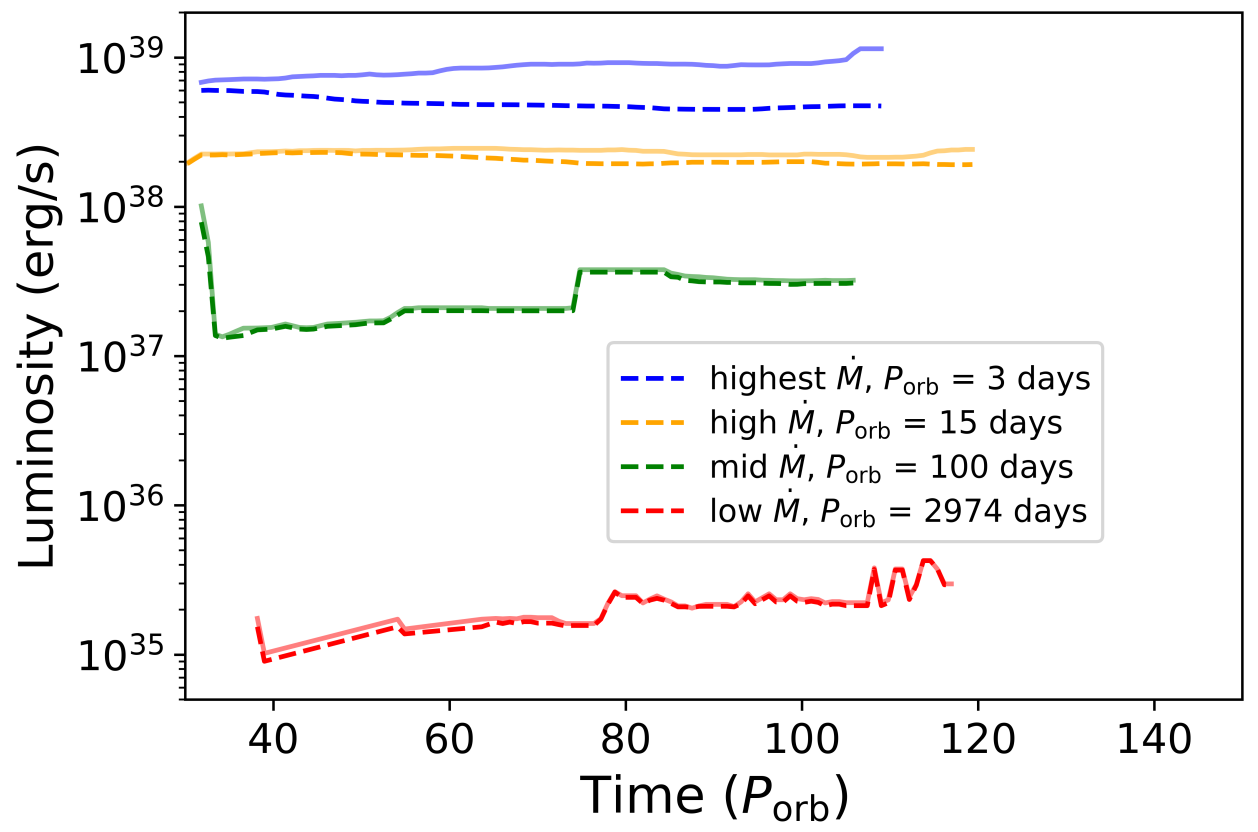

**Figure 5.** The cooling luminosity calculated within the disk around $M_a$ (dashed lines, Equation (12), and the entire computational domain (solid lines, Equation (13)). Curves are smoothed using a rolling average, and values are not shown at $t \lesssim 40$ because the MT rate has not ramped up to its steady-state value.

The increase of luminosity as a function of MT rate can be understood through an increase in accretion power from the disk surrounding $M_a$. Note, however, that we do not resolve the surface of the accretor, as it is represented by a point mass with a softening potential of length $\epsilon = 0.05a$ (Section 2). When we assume that $L_{\rm cool,disk} \sim \frac{GM_a\dot{M}_a}{r_{d,\rm eff}}$, we find that $r_{d,\rm eff}/a \approx 0.05, 0.15, 0.2, 0.3$ for our simulations of increasing MT rate (`low_mdot`, `mid_mdot`, `high_mdot`, `highest_mdot`). For the lower MT rate simulations, $r_{d,\rm eff}$ is closer to the softening radius, suggesting that the cooling power is mostly coming from the inner edge of $M_a$'s disk. For the higher MT rate simulations, $r_{d,\rm eff}$ is closer to the Roche radius of $M_a$. $\dot{M}_a$ is much less than $\dot{M}_{\rm donor}$ in the case of the `highest_mdot` and `high_mdot` simulations, which show highly non-conservative MT.

Again, while our simulations include shock heating of the gas, they do *not* include viscous heating associated with accretion. Therefore the gas is likely to experience additional heating that, in equilibrium, also increases the cooling luminosity. Accretion onto a compact object would generate heating much larger than the shock heating occurring in the accretor's outer disk. The cooling luminosities we have calculated are therefore likely lower bounds in such cases.

The gas temperature $T$, defined by Equation (6), also varies greatly between simulations. Figure 6 plots $T$ in the equatorial plane, at representative snapshots. For all simulations, $T$ in $M_a$'s disk is hotter than in the rest of the domain, due to shock heating and slower cooling. With decreasing MT rate, $T$ is lower everywhere (note the difference in colorbar scales between subplots). For the `low_mdot` simulation, $T$ outside $M_a$'s disk reaches close to a temperature floor, which can be outside our opacity tables that extend down to 1000 K. However, because there is only a small amount of gas outflowing in this case, its cooling is not important. With the differences in $T$ within the simulation domain, and between simulations, as well as the vastly different density scale between simulations (Figure 1), we expect gas pressure and radiation pressure to respectively dominate in different regions and for different simulations (see Appendix A for further discussion).

We also estimate the effective temperature $T_{\rm eff}$ of the emitted radiation by computing azimuthally averaged effective temperatures inside the disk around $M_a$ and also in the circumbinary gas, and estimating that

$$\sigma_{\rm SB} T_{\rm eff}^4 = \int \dot{E}_{\rm cool}\, dz. \tag{14}$$

We perform this integral over $z$ for cylinders of radius $R_{\rm disk}$ centered around $M_a$

$$R_{\rm disk} \equiv ((x-1)^2 + y^2)^{1/2} \tag{15}$$

and cylinders of radius $R_{\rm out}$ centered around the origin

$$R_{\rm out} \equiv (x^2 + y^2)^{1/2}. \tag{16}$$

To generate our quoted values, we then take the average of $T_{\rm eff}$ around the cylindrical axis (an average over $\phi$ in the origin-centered case). Note that Equation (14) approximately reduces to

$$T_{\rm eff,est}^4 = T^4\left(\tau_z + \frac{1}{\tau_z}\right)^{-1} \tag{17}$$

given our cooling form.

Figure 7 plots time-averaged values of $T_{\rm eff}$ both in $M_a$'s disk and CBO, versus $R_{\rm disk}$ and $R_{\rm out}$ respectively. The typical values of $T_{\rm eff}$ decrease with decreasing MT rate - unsurprisingly given that higher MT rates correspond to a higher cooling luminosity (Figure 5). As a function of increasing cylindrical radius $R_{\rm disk}$ around $M_a$ or $R_{\rm out}$ around the binary, Figure 7 demonstrates that $T_{\rm eff}$ typically decreases monotonically, which is consistent with the equatorial plots of Figure 6.

The value of $T_{\rm eff}$ reaches very low values in the circumbinary gas for the `low_mdot` simulation, which should not be trusted. In the case of such low density gas, our assumptions related to radiative cooling adopted in the simulation setup (Section 2.1), e.g., that there is an equilibrium between gas and radiation, might break down. Since the circumbinary densities and optical depths are very low, any emission would likely occur via lines (e.g., molecular lines such as CO). Dust formation may also occur and change the opacity of the gas.

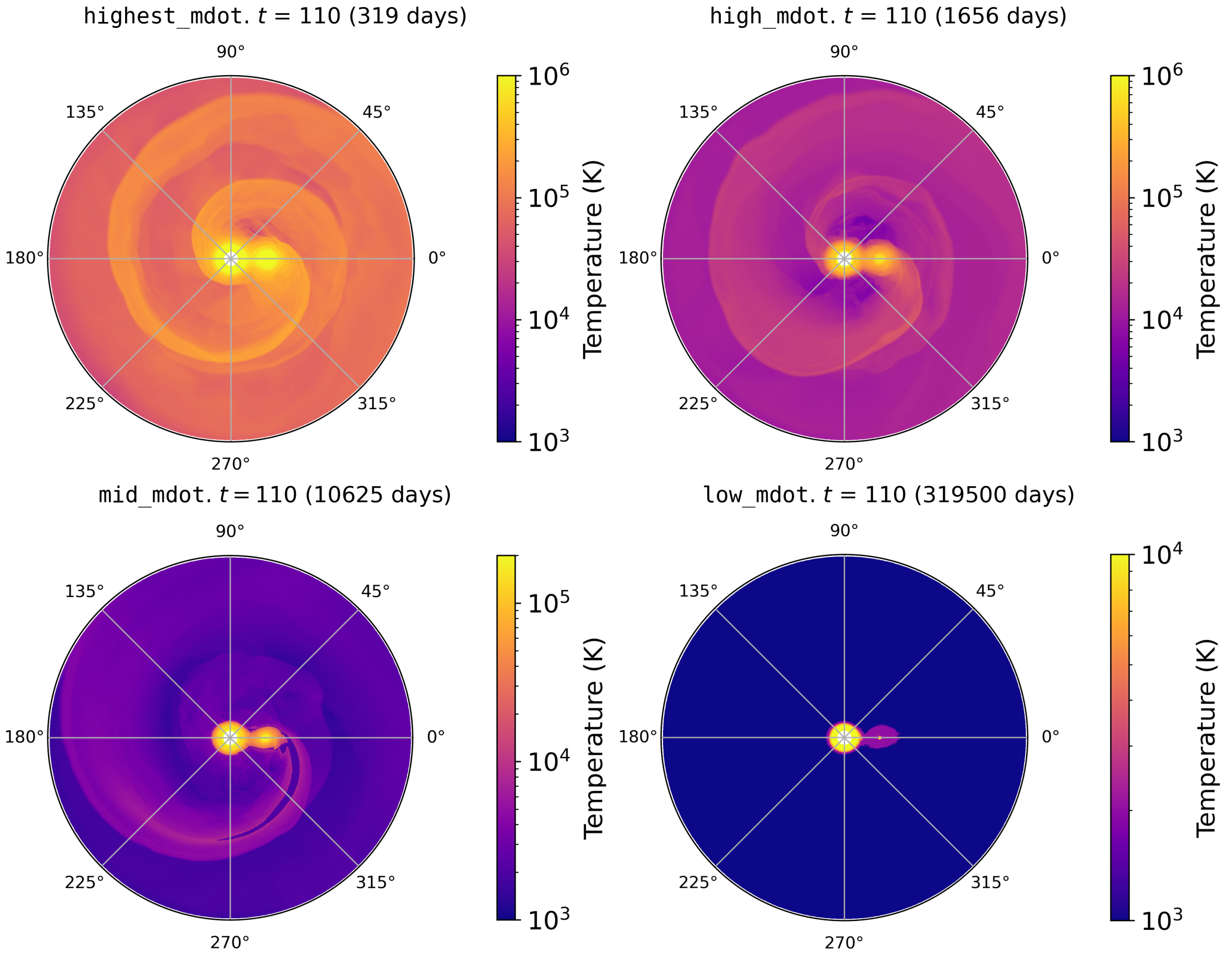

**Figure 6.** The gas temperature $T$ in the equatorial plane for all simulations. The snapshots, representative of the simulations once they have reached a quasi-steady state, are labeled with $t$ in both orbital periods and physical units (days). The $T$ floor is reached in the outer regions of the `mid_mdot` simulation and almost everywhere in the `low_mdot` simulation (see text for details).

As in our calculation of $L_{\rm cool}$, because our simulations do not include viscous heating, the values of $T$ and $T_{\rm eff}$ for a more realistic accretion disk around a compact object could be significantly higher than the values calculated here.

### 3.2.1. Observable Predictions

From the $T_{\rm eff}$ estimated in Section 3.2, we can estimate the observational appearance of the disk around $M_a$ and outflows we model, although making more precise predictions will require more realistic radiative transport modeling. For $T_{\rm eff} \sim 5 \times 10^4$, $10^4$, and $5 \times 10^3$ K in $M_a$'s disk of the `highest_mdot`, `high_mdot`, and `mid_mdot` simulations respectively, this roughly corresponds to peak emission wavelengths in the ultraviolet, near-ultraviolet/optical, and optical/near-infrared. For $T_{\rm eff} \sim 10^4$, $2 \times 10^3$, and $5 \times 10^2$ K in the CBO of the same simulations, this roughly corresponds to peak emission wavelengths in the near-ultraviolet/optical, near-infrared, and the mid-infrared, respectively.

The cooling luminosities associated with our simulations are $\sim 10^{39}$ erg s$^{-1}$ for our `highest_mdot` simulation and $\sim 10^{38}$ erg s$^{-1}$ for our `high_mdot` simulation. Even when just including the domain outside $M_a$'s disk, luminosities for these two higher MT rate simulations are $\sim 5 \times 10^{38}$ erg s$^{-1}$ and $\sim 2 \times 10^{37}$ erg s$^{-1}$. Our lower MT rate simulations are associated with lower luminosities, dominated by the cooling power in $M_a$'s disk For instance, the `mid_mdot` simulation has luminosities of $\sim 10^{37}$ erg s$^{-1}$ when including the entire domain, and $\sim 10^{36}$ erg s$^{-1}$ outside $M_a$'s disk.

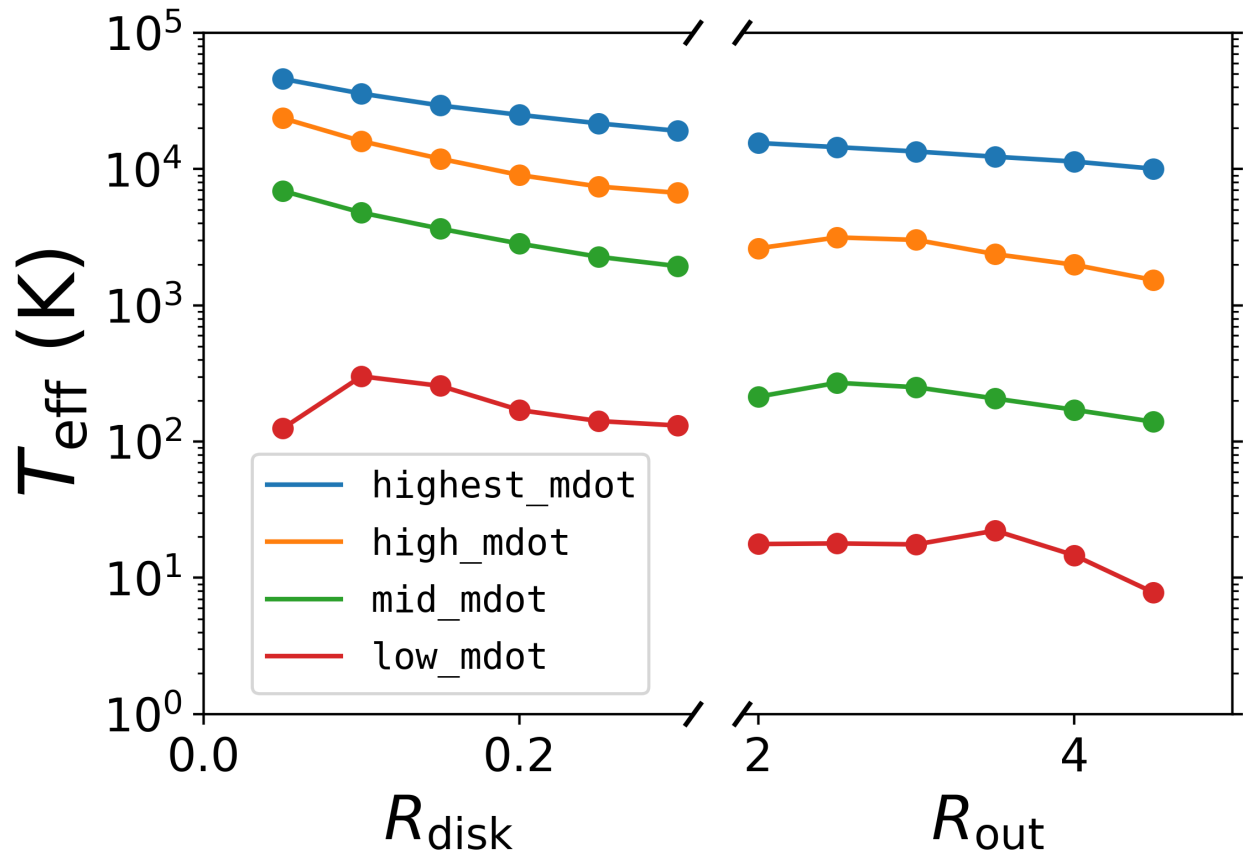

**Figure 7.** The average gas effective temperature $T_{\rm eff}$ in the disk around $M_a$ and in the circumbinary outflow. The curves shown are calculated over cylinders of constant cylindrical radius $R_{\rm disk}$ about $M_a$ and constant $R_{\rm out}$ about the origin (see text for details). $T_{\rm eff}$ values are averaged starting at $t = 50$ because the MT rate has not reached a steady-state value and the disks/outflow have not yet formed at early times.

Despite these predictions, there are multiple radiation processes that we have neglected. Although the cooling luminosity of $M_a$'s disk tends to peak near its center (Section 3.2), we have neglected a detailed treatment of accretion power onto the surface of $M_a$, especially if is a compact object like a neutron star. In such a case, a fast wind will likely be launched due to viscous accretion, which will emit X-ray and UV photons. These photons will be reprocessed to longer wavelengths by the CBO—in the infrared if dust is formed (Lu et al. 2023).

However, the production of dust in the outflow is unclear. Temperatures below around 1500 K are necessary, though not sufficient, to cause dust condensation. In the `mid_mdot` simulation, such low temperatures are reached in parts of the domain above the equatorial plane, and in the `low_mdot` simulation such temperatures are reached almost everywhere outside $M_a$'s disk, which is kept hotter by an imposed temperature floor (Figure 6). However, our low temperature estimates may not be accurate and are likely may be more associated with ambient gas rather than with the outflows we primarily seek to model. A more important question is whether dust forms somewhere in the equatorial plane of our simulations, where the outflow is strongest, especially for our higher MT simulations. Although $T$ decreases moving radially outward (Figure 7), our simulation domain does not extend far enough from the binary to capture its ultimate asymptotic temperature. Pejcha et al. (2016) shows that, for a MT rate out of L2 of $10^{-5} M_\odot$ yr$^{-1}$, the gas temperature does not drop low enough to favor dust formation until $r = 8a$. This MT rate is intermediate to our `low_mdot` and `mid_mdot` simulations, and the gas temperatures in Pejcha et al. (2016) are approximately similar to ours where the simulation domains overlap. However our domain does not extend far enough to reach lower temperatures (especially in the equatorial plane), preventing us from making more detailed predictions of dust formation.

### 3.3. Energetics, Velocities, and Angular Momentum Losses

We compute the specific AM carried away by the outflowing gas. The specific AM (along $\hat{z}$) of the L2 point is given by $h_{\rm L2} = \Omega (r_{\rm L2} - r_{\rm com})^2$, where $r_{\rm L2}$ and $r_{\rm com}$ are the radial coordinates of L2 and the COM. In units of $\sqrt{GM_{\rm tot} a}$, $h_{\rm L2} = 1.56$ for the $q = 0.5$ binaries that we simulate.

The specific AM $\boldsymbol{h}$ of an arbitrary fluid element, with coordinate $\boldsymbol{r}$ and velocity $\boldsymbol{v}$ in the simulation frame, is given by

$$\vec{h} = (\vec{r} - \vec{r}_{\rm com}) \times (\vec{v} + \vec{\Omega} \times (\vec{r} - \vec{r}_{\rm com})) \tag{18}$$

where $\boldsymbol{r}_{\rm com}$ is the coordinate of the binary's COM (in the co-rotating frame) and $\boldsymbol{\Omega}$ is the orbital frequency. We are interested predominantly in the $\hat{z}$ component of the AM, as the orbital axis of the binary is along $\hat{z}$. The rate of AM outflow $\dot{L}_z$ over a surface with area element $\boldsymbol{dA}$ is given by

$$\dot{L}_z = \int_A h_z \rho \boldsymbol{v} \cdot \boldsymbol{dA}. \tag{19}$$

Similarly, the rate of mass flow through the same surface can be written as

$$\dot{M} = \int_A \rho \boldsymbol{v} \cdot \boldsymbol{dA} \tag{20}$$

and the average specific AM $h_{\rm loss}$ of the outflowing material through this surface is the ratio of these two integrals, i.e.,

$$h_{\rm loss} = \frac{\dot{L}_z}{\dot{M}}. \tag{21}$$

As in Paper I, we find that the value of $h_{\rm loss}$ asymptotes with distance from the COM to an approximately constant value. We calculate $h_{\rm loss}$ over spheres of constant $|\boldsymbol{r} - \boldsymbol{r}_{\rm com}| \approx 3, 4,$ and $\approx 4.65$, where 4.65 is the largest COM-centered sphere that fits in the domain. Figure 8 plots the values of $h_{\rm loss}$, in units of $h_{\rm L2}$. In all cases, $h_{\rm loss} \sim h_{\rm L2}$, meaning that the outflow efficiently extracts AM from the binary - note that $h_{\rm L2}$ is much larger than the specific AM of the accretor, $h_{\rm M_a}$. For the `highest_mdot` and `high_mdot` simulations, $h_{\rm loss}/h_{\rm L2}$ is slightly less than unity, as in Paper I, whereas it is slightly larger than unity in the `low_mdot` and `mid_mdot` cases. However, even though AM may be mildly more efficiently extracted at these lower MT rates, far less material is actually flowing out

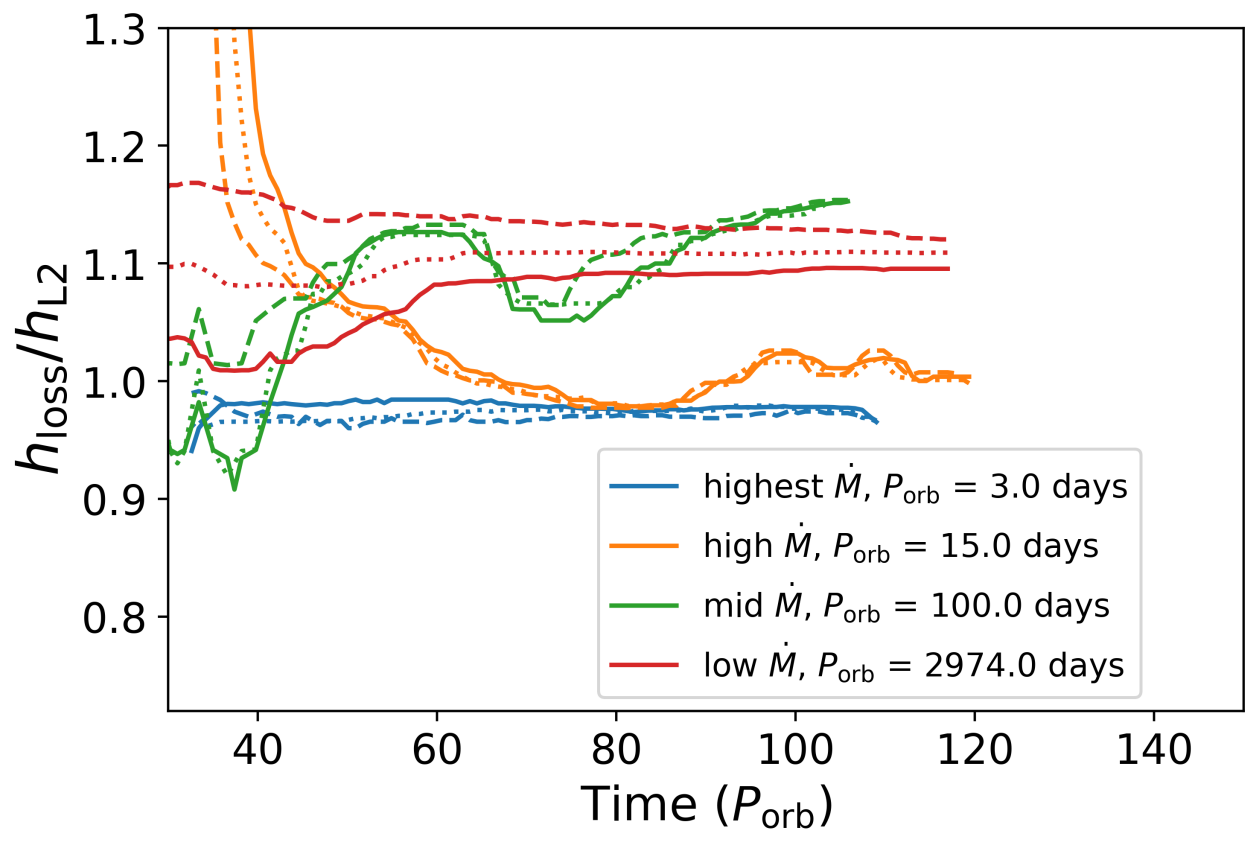

**Figure 8.** The specific AM of outflowing material $h_{\rm loss}$, in units of $h_{\rm L2}$, versus time in orbital periods. $h_{\rm loss}$ is calculated over spheres centered at the COM, of radius 3, 4, and 4.65 (dashed, dotted, solid lines) and all curves are smoothed using a rolling average.

through L2. These competing affects are discussed further in Section 3.4.

We also compute the velocity and energy of the outflowing gas. Velocities relative to the center of mass are given by $\boldsymbol{v}_{\rm inertial} = \boldsymbol{v} + \boldsymbol{\Omega} \times (\boldsymbol{r} - \boldsymbol{r}_{\rm com})$. The radial component $v_{r,\rm inertial}$ is calculated as

$$v_{r,\rm inertial} = \boldsymbol{v}_{\rm inertial} \cdot \frac{\boldsymbol{r} - \boldsymbol{r}_{\rm com}}{|\boldsymbol{r} - \boldsymbol{r}_{\rm com}|}. \quad (22)$$

The Bernoulli parameter per unit mass, $B$, is calculated as

$$B = \frac{|\boldsymbol{v}_{\rm inertial}|^2}{2} + u_{\rm int} + \Phi_{\rm grav} + \frac{P}{\rho}. \quad (23)$$

where $\Phi_{\rm grav} = -\frac{M_{\rm d}}{|\boldsymbol{r}|} - \frac{M_{\rm a}}{|\boldsymbol{r} - a\hat{x}|}$. We define the mass-weighted angle-averaged Bernoulli parameter $\bar{B}$ as

$$\bar{B} = \frac{\int \rho B dA}{\int \rho dA}, \quad (24)$$

where the integral is performed over spheres at constant $|\boldsymbol{r} - \boldsymbol{r}_{\rm com}|$.

The value of $\bar{B}$ asymptotes with radius, as in Paper I. Similar to the calculation above regarding $h_{\rm loss}$, we calculate $\bar{B}$ over spheres of constant $|\boldsymbol{r} - \boldsymbol{r}_{\rm com}|$. These values are plotted in the left hand panel of Figure 9. The `low_mdot` and `mid_mdot` simulations have near zero or slightly positive asymptotic $\bar{B}$, meaning a large fraction of the outflow is formally unbound, whereas the `highest_mdot` and `high_mdot` simulations are bound, similar to Paper I. However, although we do include radiative cooling, we do not include any momentum deposition due to irradiation of the circumbinary gas by the stars or by $M_a$'s disk that may help drive the gas outwards.

We also compared the energetics of the outflows to ballistic trajectories launched in corotation with L2, which tend to have positive asymptotic total energies and are described in more detail in Paper I. The outflows of the `highest_mdot` and `high_mdot` simulations feature asymptotic energies that are lower (more bound) compared to the ballistic trajectories, analogous to our Paper I results. In contrast, the outflows of the `low_mdot` and `mid_mdot` simulations have slightly positive energies, similar to the ballistic trajectories.

We calculate a characteristic radial velocity $\bar{v}'_{r,\rm inertial}$ using the radial inertial velocity $v_{r,\rm inertial}$ (Equation (22)),

$$\bar{v}'_{r,\rm inertial} = \frac{\int_{r=3}^{r=r_{\rm out}} \rho v_{r,\rm inertial} dV}{\int_{r=3}^{r=r_{\rm out}} \rho dV} \quad (25)$$

where we have performed a full volume integral for the average because, due to the spiral structure and shocks in the outflow, $v_{r,\rm inertial}$ can show a wave pattern that oscillates about an average value. The values of $\bar{v}'_{r,\rm inertial}$ are plotted in the right hand panel of Figure 9. In units of the orbital velocity $v_{\rm orb}$, there is a trend that the `highest_mdot` simulation has a low $\bar{v}'_{r,\rm inertial}$ (10% $v_{\rm orb}$), the `high_mdot` simulation a somewhat higher value (20% $v_{\rm orb}$), and the lower MT simulations a substantially higher value (40%–50% $v_{\rm orb}$). This is consistent with decreasing MT rate corresponding to a higher specific energy and Bernoulli parameter (left panel). In cgs units, the trend is different due to lower MT rate simulations having a lower $v_{\rm orb}$. The `low_mdot` simulation has the lowest outflow velocity at about 15 km/s, whereas the other 3 simulations have velocities clustered between 40–60 km s$^{-1}$.

Pejcha et al. (2016) performed SPH simulations, where material was initialized near L2 in co-rotation with the binary's orbit, and the outflow was tracked far from the binary to $r = 50a$. They found that that the asymptotic velocity of the outflow is always proportional to $v_{\rm orb}$. In contrast, we find that the average radial velocity in units of the escape speed, $\bar{v}'_{r,\rm inertial}/v_{\rm orb}$ tends to increase with decreasing MT rate. In addition, our outflows tend to be bound for `highest_mdot` and the `high_mdot` simulations, whereas the outflows in Pejcha et al. (2016) are unbound for the same mass ratio.

These discrepancies may be understood through MacLeod et al. (2018b) and our Paper I, which found that strong outflows through L2 tend to be initialized not at co-rotation, and are therefore more bound compared to the assumptions of Pejcha et al. (2016). Our higher MT rate simulations are in this regime. In contrast, the fact that $\bar{v}'_{r,\rm inertial}/v_{\rm orb}$ reaches a roughly constant ratio for our lower MT rate simulations seems to imply that they are in the regime consistent with Pejcha et al. (2016), although the value is a bit higher than calculated in Pejcha et al. (2016), where the asymptotic radial velocity is 0.2–0.35 $v_{\rm orb}$.

Overall, our results detailing the properties of the outflow—AM, velocity, and energy—are similar to that of Paper I, which included no radiative cooling, in the case of our

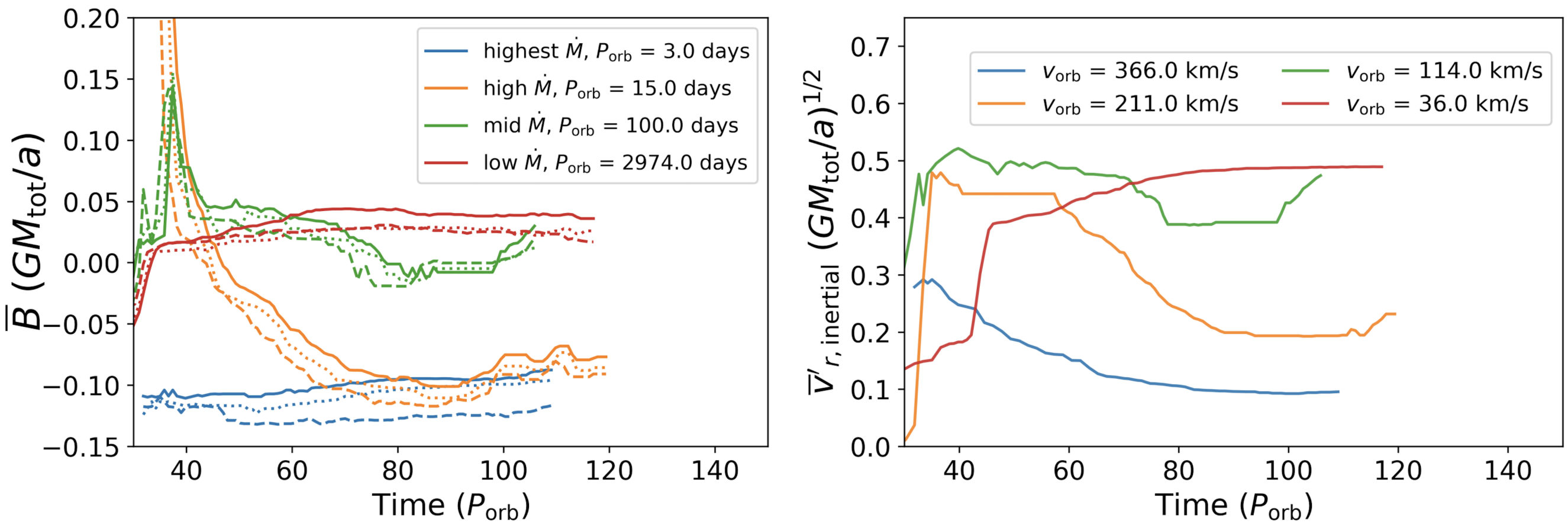

**Figure 9.** Properties related to the energy and velocity of outflowing gas. Curves are time-smoothed using a rolling average. Left: the Bernoulli parameter $\overline{B}$, in units of $GM_{tot}/a$, for COM-centered spheres of $|\boldsymbol{r} - \boldsymbol{r}_{com}| = 3$, 4 and $r_{max} \approx 4.65$, in units of $a = 1$ (dashed, dotted, solid lines)). Right: the volume-averaged radial velocity of the outflow, $\overline{v}'_{r,\mathrm{inertial}}$ (Equation (25)), in units of the orbital velocity for each simulation, $\sqrt{GM_{tot}/a}$, whose cgs value is shown in the legend.

`highest_mdot` and `high_mdot` simulations. Indeed, we found in Paper I that our adiabatic assumptions were most valid at $\gtrsim 10^{-2} M_{\odot}\,\mathrm{yr}^{-1}$. Therefore, the results of Paper I are only valid at the very high end of MT rates, and this work delineates more accurately when the results of Paper I break down.

### 3.4. Effect on Binary Orbit

The parameters $\beta$ and $h_{loss}$, describing the fraction of the transferred mass that stays in the disk surrounding $M_a$, and the AM carried by any escaping material, are of great consequence to the evolution of the binary's orbit. Because we are focused on snapshot of stable MT, we keep the orbital separation fixed in our simulations, but here discuss how the orbit would change over long-term MT.

It is useful to express the AM losses in terms of the dimensionless quantity $\gamma$, expressing the specific AM lost, in units of the specific AM of the binary. Therefore, if $J_{orb}$ is the orbital AM,

$$\gamma = \frac{h_{loss} M_{tot}}{J_{orb}} \tag{26}$$

Table 2 summarizes the approximate values of $\beta$ and $\gamma$ appropriate to the steady-state behavior of our simulations (see also Figure 3). Assuming that a fraction $\beta$ accretes onto $M_a$, which can break down if $M_a$ is a compact object and looses mass through an inner accretion disk wind, the orbital evolution of the binary then can be expressed as[4]

$$\frac{\dot{a}}{a} = -2\frac{\dot{M}_1}{M_d}\left(1 - \beta\frac{M_d}{M_a} - (1-\beta)\left(\gamma + \frac{1}{2}\right)\frac{M_d}{M_d + M_a}\right). \tag{27}$$

[4] For a derivation, see Chapter 7 of O. Pols' binary evolution notes, https://www.astro.ru.nl/~onnop/education/binaries_utrecht_notes/.

**Table 2**
For Our Grid of MT Simulations, the Fraction of Mass Retained in $M_a$'s Disk $\beta$ (Equation (9)) and Outgoing AM Parameters $h_{loss}$, in Units of $h_{L2}$, and $\gamma$ (Equations (21) and (26))

| Simulation | $\beta$ | $h_{loss}/h_{L2}$ | $\gamma$ | $\zeta_{RL}$ |
|---|---|---|---|---|
| `highest_mdot` | 0.05 | 0.95 | 6.7 | 7.9 |
| `high_mdot` | 0.4 | 1.0 | 7.0 | 6.0 |
| `mid_mdot` | 0.9 | 1.1 | 7.7 | 3.2 |
| `low_mdot` | 0.96 | 1.1 | 7.7 | 2.8 |

**Note.** $\zeta_{RL}$ is the mass–radius exponent of the donor's Roche radius (Equation (28)), with higher values generally making unstable MT more likely. The values quoted are averaged over the steady-state behavior of the simulations.

The last term on the right hand side of Equation (27) represents orbital AM lost due to outflowing gas. Even though $\gamma$ is high for the `mid_mdot` and `low_mdot` simulations, this term is close to 0 because $\beta$ is close to 1. Therefore, the additional AM loss and the effect on $\dot{a}$ is minimal in these two cases. In contrast, the `high_mdot` and `highest_mdot` simulations are associated with high values of $\gamma$ and low values of $\beta$, implying that this term will be large and $\dot{a}$ will be negative (as the term in parenthesis will be negative, but $\dot{M}_1$ is also negative). Although the orbit is expected to shrink in these cases, the stability of MT is determined also by the response of the donor star. This is beyond the scope of this work, as we drive MT through via heat injection and do not accurately model the donor's structural response.

However, we can estimate the value of $\zeta_{RL}$, the mass–radius exponent for the Roche radius of the donor

$$\zeta_{RL} = \frac{\partial \ln R_L}{\partial \ln M_d} \tag{28}$$

where $R_L$ is the Roche radius of the donor star. If $\zeta_{\rm RL}$ is large compared to the exponent representing the response of the donor's radius $R_*$ to mass loss, $\zeta_*$

$$\zeta_* = \frac{\partial \ln R_*}{\partial \ln M_d} \tag{29}$$

the overfill factor of the donor will increase and MT is likely to be unstable. We calculate $\zeta_{\rm RL}$ following the procedure in Soberman et al. (1997), and expressing $\zeta_{\rm RL}$ as

$$\zeta_{\rm RL} = \frac{\partial \ln R_L/a}{\partial \ln q'}\frac{\partial \ln q'}{\partial \ln M_d} + \frac{\partial \ln a}{\partial \ln q'}\frac{\partial \ln q'}{\partial \ln M_d} \tag{30}$$

with $q' = \frac{1}{q}$, where $q'$ follows the mass ratio definition of Soberman et al. (1997). It is straightforward to show that

$$\frac{\partial \ln q'}{\partial \ln M_d} = (1 + \beta q') \tag{31}$$

and $(\partial \ln R_L/a)/(\partial \ln q')$ can be computed using the formula of Eggleton (1983) (see Equation (64) of Soberman et al. 1997). Finally, we use Equation (27) to find

$$\frac{\partial \ln a}{\partial \ln q'} = \frac{-2\left(1 - \beta q' - (1-\beta)(\gamma + \frac{1}{2})\frac{q'}{1+q'}\right)}{(1+\beta q')}. \tag{32}$$

This can be shown to be identical to the relevant equation in Table 5 of Soberman et al. (1997) when mass loss is through a circumbinary ring (there appears to be a typo in Equation B13 of Soberman et al. 1997, where there is a missing factor of their $\epsilon$ in order to agree with Table 5). Also note that our $\gamma$ must be multiplied by a factor of $\frac{1/q}{(1+1/q)^2}$ to align with the definition of $\gamma$ in Soberman et al. (1997).

The values of $\zeta_{\rm RL}$ calculated for our simulations are given in Table 2. We find that $\zeta_{\rm RL}$ is large, between 6–8, for the `highest_mdot` and `high_mdot` simulations, such that MT is likely to be unstable unless $\zeta_*$ is similarly large, which would correspond to a star that contracts very rapidly in response to MT. This means that a star with a radiative envelope, with $\zeta_* \gg 1$, may still undergo stable MT in these cases. For the two lower MT rate simulations, $\zeta_{\rm RL}$ reaches a value of approximately 3. For convective-envelope stars, with $\zeta_* < 0$, MT is likely unstable for the $q$ that we consider.

## 4. Conclusion

Episodes of rapid MT ($\gtrsim 10^{-4} M_\odot\,{\rm yr}^{-1}$) occur in many binaries containing a massive star, potentially leading to non-conservative MT and CBOs. We simulated stable MT between a 10 $M_\odot$ donor star and an 5 $M_\odot$ unresolved accretor, probing MT rates between $\sim 10^{-5}$ and $\sim 10^{-1} M_\odot\,{\rm yr}^{-1}$ by varying the orbital separation of the binary (Table 1). Our 3D simulations are performed using the PLUTO hydrodynamic code. They resolve the flow of mass from the donor, into a disk around the accretor, through the outer Lagrange point (L2), and into a CBO. Radiative cooling is included at every grid cell, using tabulated opacities and an approximate treatment of photon diffusion perpendicular to the orbital plane, where the density is lowest. Our idealized simulations do not include detailed radiative transport or radiative forces.

The main results of our work are:

1. Mass from the disk around the accretor tends to outflow in a stream originating near L2, forming an equatorially concentrated CBO (Figure 1). The fraction of gas staying in the accretor's disk, versus outflowing near L2, is expressed with the parameter $\beta$, where $\beta = 1$ implies conservative MT. We find that $\beta$ decreases strongly with increasing MT rate, meaning more material outflows, and in particular rises sharply between $\sim 10^{-3}$ and $\sim 10^{-2} M_\odot\,{\rm yr}^{-1}$ (Figure 3), approximately where the accretion luminosity of the outer edge of the accretor's disk becomes super-Eddington.
2. Across a wide range of MT rates, the outflowing gas has a specific AM $h_{\rm loss}$ that asymptotes with radius to a value approximately equal to that of the L2 point, $h_{\rm L2}$ (Figure 8) and much larger than the specific AM of the accretor. This means that, in the case of strongly non-conservative MT and significant outflows, AM will be efficiently extracted from the binary, which will have important consequences for orbital evolution.
3. The outflow reaches an asymptotic radial velocity that increases from 10% the orbital velocity to 50% the orbital velocity as the MT rate decreases from $\sim 10^{-1}$ to $\sim 10^{-3} M_\odot\,{\rm yr}^{-1}$. In physical units, however, the radial velocity is about 30–50 km s$^{-1}$ for our simulations, except for the `low_mdot` simulation where it is lower (Figure 9).
4. The gas cooling luminosity is dominated by the accretor's disk at low MT rates, but the outflow luminosity contributes strongly at $\sim 10^{-2} M_\odot\,{\rm yr}^{-1}$ and dominates the luminosity at $\sim 10^{-1} M_\odot\,{\rm yr}^{-1}$ (Figure 5), leading to luminosities between $10^{38}$ and $10^{39}$ erg s$^{-1}$.
5. The gas temperature and $T_{\rm eff}$ decrease with decreasing MT rate (Figures 6, 7), with $T_{\rm eff}$ of the CBO $\sim 10^4$, $10^3$, $10^2$ K for MT rates of $\sim 10^{-1}$, $10^{-2}$, $10^{-3} M_\odot\,{\rm yr}^{-1}$. The $T_{\rm eff}$ of the accretor's disk is an order of magnitude higher, peaking near the disk center, though we do not resolve the inner disk where we impose a softening potential. Dust formation may occur within the outflow, especially away from the equatorial plane and for our lower MT rate simulations.

Although there have been a few prior works that have performed hydrodynamic simulations involving outflows through L2 (and/or L3) that implemented radiative cooling losses, our work differs by systematically controlling the MT rate from the donor star and tracking its effect on the CBO.

Our measurement of the MT rates where MT transitions from conservative to non-conservative, as well as the AM carried by the outflowing gas, can be incorporated into population synthesis calculations investigating the stability of MT, and orbital evolution during stable MT preceding mergers of binary compact objects (e.g., Gallegos-Garcia et al. 2021; Marchant et al. 2021). Observationally, these simulations can be used to predict the formation of material following extreme mass loss from supernova progenitors (Wu & Fuller 2022), as well as the temperature and luminosity of outflows associated with a phase of rapid MT. However, more examination of the outflow farther away from the binary is needed to investigate the formation of dust, especially in the equatorial plane of the outflow, and the outflow's ultimate properties, which likely connect to luminous red novae (Pejcha et al. 2016). In addition, our simulations were performed for only a single combination of donor and accretor masses, and for solar metallicity and composition, which should be extended to other massive binaries.

## Acknowledgments

We are grateful for support from the NSF through grant AST-2205974. and to the Gordon and Betty Moore Foundation through grant GBMF5076.

This work used the Purdue Anvil supercomputer through allocation PHY250215 from the Advanced Cyberinfrastructure Coordination Ecosystem: Services & Support (ACCESS) program, which is supported by U.S. National Science Foundation grants #2138259, #2138286, #2138307, #2137603, and #2138296.

We thank the anonymous referee for their thoughtful comments which greatly improved this work.

*Software:* PLUTO, Python, NumPy (Harris et al. 2020), Matplotlib (Hunter 2007), SciPy (Virtanen et al. 2020), Roche lobe calculator (Leahy & Leahy 2015), Roche_tidal_equilibrium[5] (Lu 2025), L2massloss.[6]

## Appendix A
## Ratio of Gas to Radiation Pressure

Figure 10 plots the ratio of gas pressure to radiation pressure in the equatorial plane, where, given $\rho$ and P, the two terms are calculated via

$$P = \frac{\rho k_b T}{\mu m_p} + \frac{a_{\rm rad} T^4}{3}. \tag{A1}$$

See Section 2.1 for more details related to the calculation. In general, regions with higher $\rho$ are more gas pressure dominated. This includes $M_a$'s disk and the CBO originating at L2 (and L3, for the `highest_mdot` and `high_mdot` simulations. Where the density is low, radiation pressure dominates (the purple voids of Figure 10). As the circumbinary gas density decreases with decreasing MT rate, likewise the simulations with decreasing MT rate are less gas pressure dominated in the circumbinary regions. However, the opposite is true in the disk around the accretor. For the `low_mdot` and `mid_mdot` simulations, the ratio of gas to radiation pressure reaches quite a high value in the center of $M_a$'s disk, where the gas has collapsed to a cool, dense disk. Note that we are using an adiabatic index of 1.4 for our EOS (Equation (1)) due to limitations in the current EOS options in PLUTO. Future work should adopt a more dynamic equation of state that varies more self-consistently with $T$ and $\rho$.

[5] https://github.com/wenbinlu/Roche_tidal_equilibrium
[6] https://github.com/wenbinlu/L2massloss

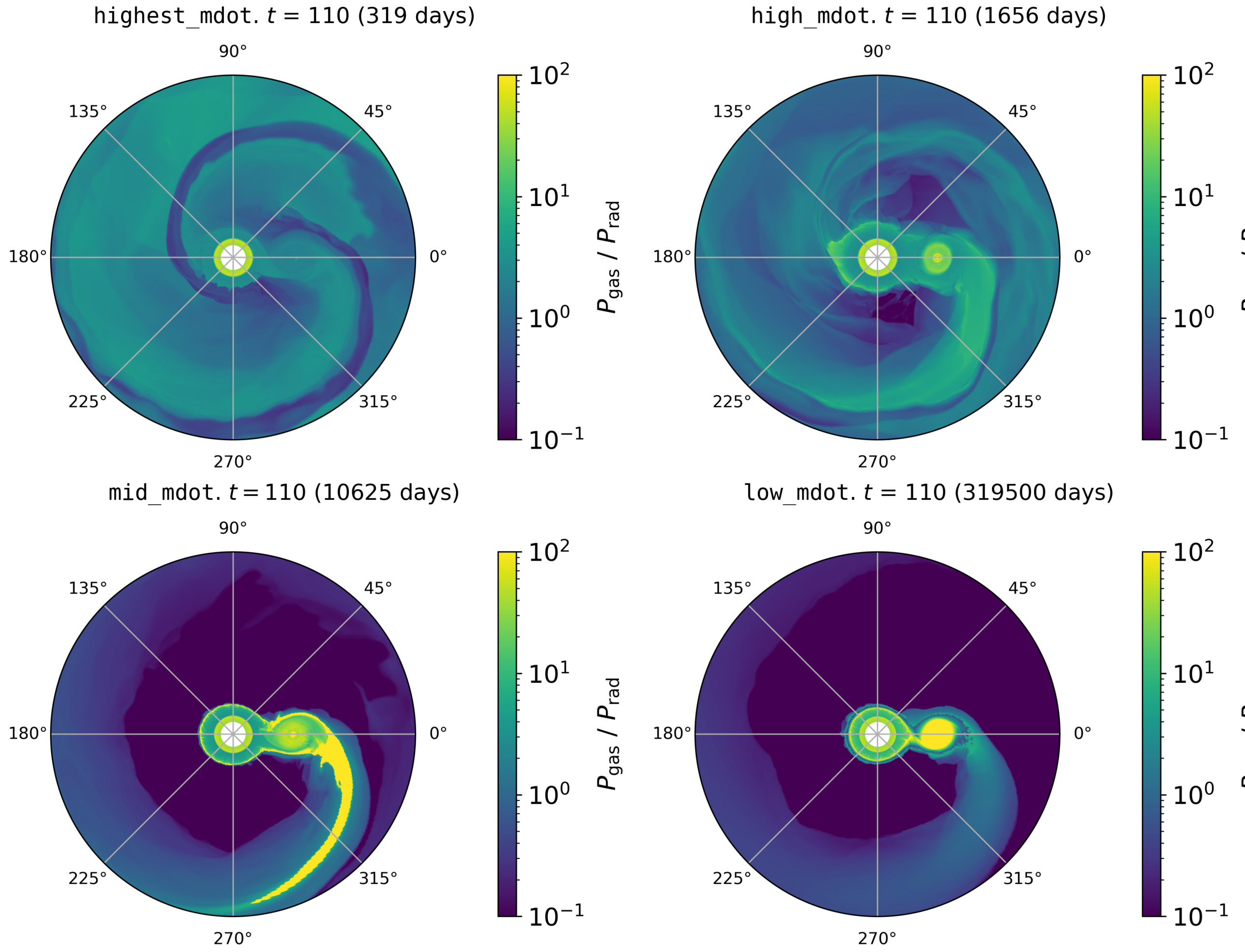

**Figure 10.** The ratio of gas pressure to radiation pressure (see Equation (A1)) in the equatorial plane of the simulations, for snapshots representative of the quasi-steady state behavior.

## Appendix B Timescales

The ratio of the cooling timescale to mass advection timescale changes significantly between our simulations of different MT rates. The cooling timescale $t_{\rm cool}$ is given by

$$t_{\rm cool} = \frac{u_{\rm int}}{|\dot{E}_{\rm cool}|}. \tag{B1}$$

The value of $t_{\rm cool}$ is estimated, with density-weighted averages, 1. within the $M_a$'s disk at several different radii surrounding $M_a$ and inside its Roche lobe; 2. at radii within the CBO, at radii centered at the origin and larger than $a$.

The advective timescale $t_{\rm adv}$ is given by

$$t_{\rm adv} = \frac{|\boldsymbol{r} - a\hat{x}|}{v_{r,M_a}} \tag{B2}$$

near $M_a$, where $v_{r,M_a}$ is the component of velocity radially outward from $M_a$. In the broader CBO surrounding both stars, we estimate $t_{\rm adv}$ as simply

$$t_{\rm adv} = \frac{|\boldsymbol{r}|}{v_r} \tag{B3}$$

where $v_r$ is the radial component of velocity relative to the origin. We similarly perform density-weighted averages of $t_{\rm adv}$ at various radii in our domain, using the appropriate definition in $M_a$'s disk or in the CBO.

For the `highest_mdot` and `high_mdot` simulations, we find that $t_{\rm cool}$ is larger than $t_{\rm adv}$ in both $M_a$'s disk and the CBO. Because the gas is optically thick everywhere, it is difficult for radiation to escape in this case. For the `mid_mdot` simulation, $t_{\rm adv}$ is smaller than $t_{\rm cool}$ in $M_a$'s disk, but is larger in the CBO. This implies that the outflow cools quickly once it leaves $M_a$'s disk. Finally, for the `low_mdot` simulation, $t_{\rm adv}$ is larger than $t_{\rm cool}$ in both $M_a$'s disk and in the CBO, meaning that material rapidly cools to near the internal energy floor (which is set via a temperature floor of 2000 K in the higher density regions such as the $M_a$'s disk).

## Appendix C
## Adopted Opacities

Figure 11 plots the tabulated value of $\kappa$ used in our work as a function of $\rho$ and $T$, as well as some representative points showing where the cells in the simulation domain are located in this phase-space for a representative snapshot. The opacity rises at $\log T \gtrsim 3.7$ due to $H^-$ opacity, and is dominated by dust for $\log T \lesssim 3.1$ (Pejcha et al. 2016).

We divide the grid domain into three representative regions: inside $M_a$'s Roche radius (cyan points); in the equatorial plane, where $\rho$ is the highest, but outside $M_a$'s Roche radius (red points); everywhere else in the domain, which are therefore points above the equatorial plane and not near $M_a$ (orange points).

The black dashed lines of Figure 11 show the floors that we impose (see also Section 2.1 for the reasoning behind our floors). The vertical line at $T = 2000$ K represents our temperature floor imposed in relatively high $\rho$ regions, whereas the curved line shows the floor on sound speed that is imposed in low density regions. The horizontal dashed line shows the density floor imposed everywhere. For the latter two lines, the floor values are the same in code units, but vary in physical units and so shift position between the two subplots.

For the high_mdot simulation shown in Figure 11, points near $M_a$ are safely away from the imposed floors. In this region, the dominant opacity appears to from free–free and bound-free transitions, although this is a rough estimate based on a Kramer's opacity law. Some points in the equatorial plane and many points above the equatorial plane reach the sound speed floor, which is essentially a density-dependent temperature floor. Points above the equatorial plane also tend to approach the density floor. For the mid_mdot simulation, most points near $M_a$ are above the temperature floor, although some approach it. Most points elsewhere in the domain approach the density and sound speed floor. The $T = 2000$ K floor is not relevant because of the low densities in these regions - were it imposed, it would lead to an artificially high internal energy and a spurious flux of mass in the outer domain. Overall, there is a progression where the high MT rate simulations never approach the temperature floor, the mid_mdot simulation reaches the floor only far away from $M_a$'s disk, and the low_mdot simulation reaches the floor basically everywhere.

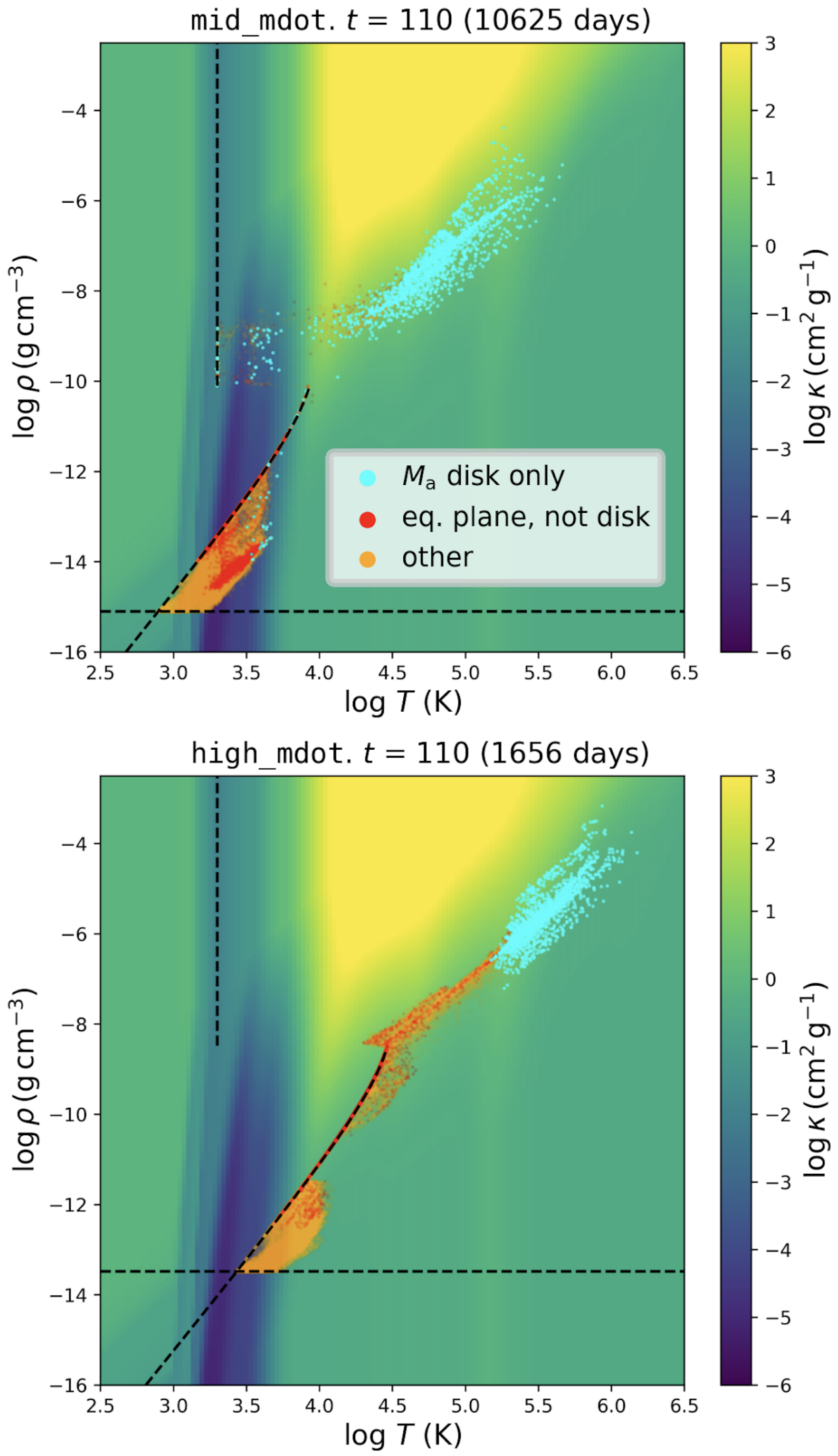

**Figure 11.** Plots of the opacity used in this work. Colored dots correspond to a sample of grid cells, corresponding to the labeled $t$, and where they lie in $\rho$, $T$ space. Cyan points are located in the Roche lobe surrounding $M_a$. Red points are in the equatorial plane, excluding any points falling into $M_a$'s Roche lobe. Orange points are sampled everywhere else in the domain, i.e., above the equatorial plane. The highest_mdot and low_mdot simulations are not shown because they are relatively similar to the subplots shown here. The black curves correspond to floors on $\rho$, $T$, and $c_s$ (see text for details). The opacity tables used (see Section 2.1) extend down to $10^3$ K, so points off the table use the value at the boundary.